\documentclass[aps,pra,twocolumn,nofootinbib,showpacs,superscriptaddress]{revtex4-1}

\usepackage{amsmath}
\usepackage{amsfonts}
\usepackage{amssymb}
\usepackage{graphicx}
\usepackage{color}
\usepackage{xcolor}

\newcommand{\reff}[1]{(\ref{#1})}
\newcommand{\dd}{\mathrm{d}}

\newcommand{\ee}{\mathrm{e}}
\newcommand{\xx}{\mathbf{x}}
\newcommand{\kk}{\mathbf{k}}
\newcommand{\vecphi}{\boldsymbol\phi}
\newcommand{\vecpi}{\boldsymbol\Pi}

\begin{document}

\title{Short-time universal scaling in an isolated quantum system after a quench}

\author{Alessio Chiocchetta}
\thanks{These authors contributed equally.}
\affiliation{SISSA --- International School for Advanced Studies and INFN, via Bonomea 265, 34136 Trieste, Italy}
\author{Marco Tavora}
\thanks{These authors contributed equally.}
\affiliation{Department of Physics, New York University, 4 Washington Place, New York, NY 10003, USA}
\author{Andrea Gambassi}
\affiliation{SISSA --- International School for Advanced Studies and INFN, via Bonomea 265, 34136 Trieste, Italy}
\author{Aditi Mitra}
\affiliation{Department of Physics, New York University, 4 Washington Place, New York, NY 10003, USA}

\begin{abstract}
Renormalization-group methods provide a viable approach for investigating the emergent collective behavior of
classical and quantum statistical systems in both equilibrium and nonequilibrium conditions.
Within this approach we investigate here the dynamics of an isolated quantum system represented by a
scalar $\phi^4$ theory after a global quench of the potential close to a dynamical critical point.
{We demonstrate that, within a pre-thermal regime, the time dependence of the relevant correlations is characterized by 
a short-time universal exponent, which we calculate at the lowest order in a dimensional expansion.}
\end{abstract}



\date{\today}

\maketitle

\section{Introduction}
The nonequilibrium dynamics of isolated, strongly interacting quantum many-body systems is currently under intensive experimental and theoretical investigation
(see, e.g., Refs.~\cite{Polkovnikovrev,Lamacraft12,Yukalov11}), primarily motivated by recent advances in the physics of cold atomic gases~\cite{Bloch08}.
A natural question which arises 
in this context concerns
the eventual thermalization of these systems after a sudden change (\emph{quench}) of a control parameter.
In fact, {although} isolated systems evolve with 
unitary dynamics \cite{GreinerMandel02,GMHB-2002},
their local properties can be described, after some time, by suitable statistical ensembles \cite{Deutsch91,Srednicki94,RigolDunjko08}.
Interestingly enough, the eventual approach to a thermal state might involve intermediate \emph{pre-thermal} quasi-stationary states,
proposed theoretically \cite{Berges04} and experimentally observed \cite{Kitagawa11a,Gring12, Langen2013}. These
states appear to be related to the integrable part of the post-quench
Hamiltonian \cite{Kollar11,Kehrein08,Moeckel09,Moeckel10,Marino12,Mitra13a,Worm13,Marcuzzi13}, which alone \cite{Weiss06}
would drive the system towards a state, sometimes well described by the so-called generalized Gibbs ensemble (GGE)
\cite{Rigol07,PhysRevA.80.063619,Jaynes57, Barthel08, Goldstein14, Pozsgay14, Mierzejewski14,Wouters14,Muss14}.
Inspired by the analogy with renormalization-group (RG) flows, pre-thermalization has been ascribed to
a non-thermal unstable fixed point \cite{Berges08, Nowak11, Nowak14}
towards which the evolution of the system is attracted before crossing 
over to the eventual, stable, thermal fixed point.

While most of the properties of an isolated {many-body} system after a quench depend on its microscopic features,
some acquire a certain degree of universality
if the post-quench Hamiltonian is close to a critical point.
Examples include the density of defects \cite{Polkovnikovrev}, dynamics of correlation functions~\cite{Huse12,Mitra13a},
statistics of the work \cite{Gambassi11,GambassiSilva12,Sotiriadis13}, rephasing dynamics \cite{Torre12b}, 
dynamical phase transitions \cite{Sciolla2010,Gambassi2011,Sciolla11,Sciolla2013,Sondhi2013,Smacchia,Werner09,Schiro2010},
or the dynamics of solitons \cite{Franchini2014}.
Despite this progress, an important open issue is the possible emergence of a \emph{universal} collective behavior at \emph{macroscopic short-times} controlled by the memory of the initial state, i.e., a kind of \emph{quantum aging}.
This is known to occur for quenches in classical systems in the presence of a thermal bath \cite{Janssen89,Gambassi05,Bonart12,Marcuzzi12} and, more 
recently, for quantum impurities \cite{Hackl2009, Pletyukhov2010} or open quantum systems \cite{Gagel14,Buchhold14}. A quench introduces a ``temporal boundary'' 
by breaking the time-translational invariance (TTI) that characterizes equilibrium dynamics, causing the emergence of 
short-time universal scaling, analogous to universal short-distance 
scaling in the presence of spatial boundaries in equilibrium~\cite{Diehl86,Diehl97,Pleimling04}.
{To our knowledge, non-equilibrium dynamical scaling and aging 
have never been investigated in the absence of a thermal bath. 
In this work, we fill this gap by showing the emergence of these features 
after a quench of an \emph{isolated} quantum many-body system.}

At the lowest order in a dimensional expansion, we construct the RG equations for a wide class of isolated quantum systems after a quench, discussing the resulting flow and comparing it with the equilibrium one at a certain effective temperature $T_{\rm eff}$. {Remarkably, these RG equations are characterised by 
a stable non-Gaussian fixed point which is associated with the occurrence of a 
\emph{dynamical phase transition} (DPT)}.
Similarly to the case of classical and quantum systems in contact with thermal baths  
mentioned above, we show the appearance of universal algebraic 
laws associated with such non-thermal fixed point, which determines the temporal scaling of the relevant quantities, and which is later on destabilized by the thermalizing dynamics.

\section{The model} 
In $d$ spatial dimensions consider a system belonging to the equilibrium universality class described by the 
effective $O(N)$-symmetric Hamiltonian
\begin{equation}
\label{eq:Hamiltonian}
{H(r,u) = \int \dd^d x \left[ \frac{1}{2}\vecpi^2 + \frac{1}{2}(\nabla \vecphi)^2 + \frac{r}{2}\vecphi^2+\frac{u}{4!N}\vecphi^4\right],}
\end{equation}
{where $\vecphi = (\phi_1,\ldots,\phi_N)$ is a bosonic field with $N$ components, $\vecpi$ its conjugate momentum,
$u>0$,  and $r$ the parameter which controls the distance from the critical point.}
The system is prepared at $t<0$ in the ground state of the non-interacting Hamiltonian $H_0\equiv H(\Omega_0^2,0)$, in a highly disordered phase ($\Omega^2_0> 0$), and at
time $t=0$ the parameters are suddenly changed, resulting in the post-quench Hamiltonian $H\equiv H(r,u)$. 
The quench is performed towards a disordered or critical phase such that, in the absence of symmetry-breaking fields, the order parameter $\bar \vecphi(t) \equiv \langle \vecphi \rangle$ vanishes during the dynamics.
$H$ for $u=0$   as well as   $H_0$ can be diagonalized in momentum space in terms of two sets of creation/annihilation operators with dispersion relation  $\omega_k(r) = \sqrt{k^2 + r} \equiv \omega_k$ and $\omega_k(\Omega^2_0) \equiv \omega_k^0$, respectively, where $k$ is the modulus of the momentum.
By requiring the continuity of $\vecphi$ and $\vecpi$ during the quench $\Omega^2_0\to r$,
these two sets of operators are related by a Bogoliubov transformation \cite{Mahan}. 
The relevant two-time correlation functions which characterize the ensuing dynamics  
are the retarded and the Keldysh nonequilibrium Green's functions \cite{Kamenevbook}, defined respectively as
$iG_{\alpha\beta,R}(1,2) = \vartheta(t_1-t_2) \langle \left[\phi_\alpha(1),\phi_\beta(2)\right]\rangle$ [where $\vartheta(t>0)=1$ and $\vartheta(t<0)=0$] and $ iG_{\alpha\beta,K}(1,2) = \langle\left\{\phi_\alpha(1),\phi_\beta(2)\right\}\rangle$, with
$n \equiv (\xx_n,t_n)$ and $\alpha$, $\beta$ specifying the components of the field. These functions are non-zero only for $\alpha=\beta$ and they do not depend on $\alpha$ in the symmetric phase, i.e., $G_{\alpha\beta,K/R}= \delta_{\alpha\beta}G_{K/R}$. Their Fourier transforms read:
\begin{align}
&G_R(k,t_1,t_2) = -\vartheta(t_-) \frac{\sin(\omega_k t_-)}{\omega_k},  \label{eq:GR}\\
&iG_K(k,t_1,t_2) =  \frac{K_+ \cos(\omega_kt_-)+ K_- \cos(\omega_k t_+)}{\omega_k}, \label{eq:GK}
\end{align}
for $u=0$, where $t_\pm = t_1\pm t_2$ and $K_\pm (k)  = (\omega_k/\omega_k^0\pm \omega^0_k/\omega_k)/2$.
Note that $G_K$ (but not $G_R$) depends on the pre-quench state  
and is not TTI.
Hereafter we primarily focus on the case $\Omega_0\gg \Lambda$, where $\Lambda$ is the momentum cutoff introduced further below; on a lattice, this implies that the spatial correlation length in the initial state is smaller than the lattice spacing. As the RG fixed-point value of $r$ turns out to be of order $\Lambda^2$ (see further below), this case actually corresponds to $\Omega_0^2 \gg r$ and therefore to a \emph{deep quench} of the coefficient of $\vecphi^2$ in Eq.~\eqref{eq:Hamiltonian}.
The stationary part $\Omega_0\cos(\omega_kt_-)/(2\omega_k^2)$ of $iG_K$ turns out to have the same form as in equilibrium~\cite{Mahan} 
at a high temperature $T=\Omega_0/4 \gg \Lambda$ (see also Refs.~\cite{Calabrese07,PhysRevB.89.134307}).
A similar conclusion holds for the (non-thermal) occupation number $n_k$ of the post-quench momenta,
which is approximately thermal for $k \ll \Omega_0$.
Accordingly, the behavior of the system after the quench is expected to bear some similarities to the equilibrium one at temperature $T$.
Depending on $d$ and $N$, the latter encompasses an order-disorder transition at 
$r=r^*_{\rm eq}(T)$ \cite{Sachdev11,Sondhi97,Fisher88}
which displays the critical properties of a classical system in $d+1$ spatial dimensions for $T=0$, while those of a classical
system in $d$ dimensions for $T>0$ because, in this case, the additional dimension has a finite extent $T^{-1}$.
On this basis, after the quench, one heuristically expects a collective behavior to emerge at some value $r^*(\Omega_0)$ of $r$,
as in a $d+1$-dimensional film of thickness $\sim \Omega_0^{-1}$.
In addition,
the non-stationary part $-\Omega_0 \cos(\omega_kt_+)/(2\omega_k^2)$ of $i G_K$ (absent in equilibrium)
turns out to be responsible for the short-time universal scaling behavior discussed below.

{The case of a quench which does not affect $u$, i.e., which occurs from the ground state of $H(r_0,u)$ to $H(r,u)$,
was studied within the mean-field approximation in Ref.~\cite{Gambassi2011} 
and in the exactly solvable limit $N\to \infty$ in Refs.~\cite{Sondhi2013,Sciolla2013,Smacchia}.
Quite generically it was shown that, upon crossing a line in the $(r_0,r)$-plane (at fixed $u$), the system undergoes a dynamical transition signaled by a qualitative change in the time evolution of the mean order parameter $\bar\vecphi$. In particular, starting from a disordered initial state with $\bar\vecphi=0$ (i.e., $r_0>0$), this transition occurs at a certain 
$r= r^*<0$, below which the system undergoes coarsening. 
Although the quench protocol considered here involves a vanishing pre-quench $u$, a non-vanishing $u$ 
solely affects the effective value of $r_0=\Omega_0^2$. 
Accordingly, we expect that the DPT associated with the RG fixed point $Q_{\rm dy}$ discussed further below and emerging after the quench is closely  
related to the DPT discussed in Refs.~\cite{Gambassi2011,Sondhi2013,Sciolla2013,Smacchia}.
Indeed, the critical exponent $\nu$ which describes the RG flow around $Q_{\rm dy}$
agrees, up to the first order in the dimensional expansion 
and for $N\to\infty$, with the exact result found in Ref.~\cite{Smacchia} at the dynamical transition.}  

\section{Renormalization-group flow}
In order to highlight the dynamical scaling  
after the quench and to account for the effects of non-Gaussian fluctuations, we study perturbatively the RG flow of the relevant couplings \cite{Kamenevbook}.  In particular, from the Schwinger-Keldysh
action associated with $H$ in Eq.~\eqref{eq:Hamiltonian} we determine the effective action for the ``slow'' modes
by integrating those with a wavevector $k$ within a shell of infinitesimal thickness just below the cutoff $\Lambda$.
Subsequently, spatial coordinates, time, and fields are rescaled in order to restore the initial cutoff $\Lambda$:
from the resulting coupling constant one infers the RG equations~\cite{Wilson74,Fisher98}.
An analogous procedure was recently carried out for a quench in $d=1$ \cite{Mitra12b,Mitra13a}, for driven quantum systems in $d>1$ (see, e.g., Refs.~\cite{Mitra08a,Sengupta14}), and for quantum impurities (see, e.g., Refs.~\cite{Hackl2009, Pletyukhov2010}). At one loop and for times larger than the microscopic time $\simeq \Lambda^{-1}$ (before which the dynamics is non-universal), the resulting RG equations read (see Appendix~\ref{app:app1})
\begin{subequations}
\label{eq:RG12}
\begin{align}
\frac{\dd r}{\dd \ell} &= 2 r + a_d \frac{N+2}{24N} u \Lambda^d \frac{2\Lambda^2 + r + \Omega_0^2}{(\Lambda^2 + r)\sqrt{\Lambda^2 + \Omega_0^2}}+ {\cal O}(u^2), \label{eq:RG1} \\
\frac{\dd u}{\dd \ell} &
= (d_c-d) u -   a_d\frac{N+8}{24N} u^2 \Lambda^{d-4} \sqrt{\Lambda^2 + \Omega_0^2} +  {\cal O}(u^3),
\label{eq:RG2}
\end{align}
\end{subequations}
where $a_d = 2/[(4\pi)^{d/2}\Gamma(d/2)]$, $d_c$ is the upper critical dimensionality discussed below, and $\ell > 0$ is the flow parameter which rescales coordinates and times as $(x,t)\mapsto (\ee^{-\ell} x, \ee^{-\ell}t)$.
%
%
%
%
\begin{figure}[t!]
\begin{center}
\includegraphics[trim = 0mm 70mm 0mm 60mm, width=0.48\textwidth]{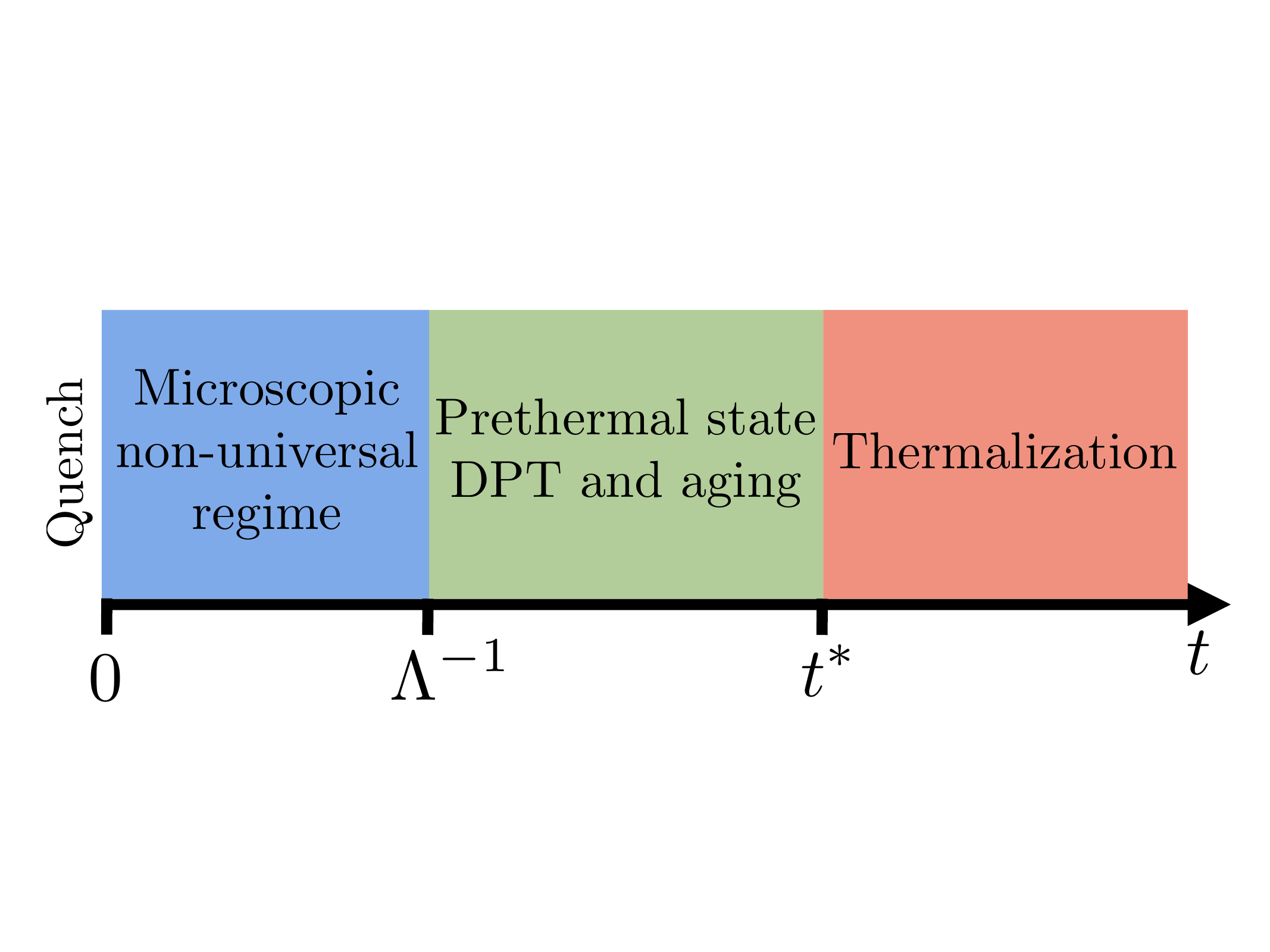}
\caption{(Color online) A schematic picture of the various temporal regimes which characterize the evolution of the system after the quench.}
\label{fig:sketch}
\end{center}
\end{figure}
%
%
{According to this scaling, the RG flow can be parameterized in terms of the time $t$ elapsed from the quench by setting $\ell = \ell_t \equiv \ln (\Lambda t)$.}
Equations \eqref{eq:RG12} are actually valid up to a typical time $t^*$ discussed later, after which thermalization may take place, according to the dynamical scenario sketched in Fig.~\ref{fig:sketch}.
For $\Omega_0\ll \Lambda$, inspection of Eq.~\reff{eq:RG2} shows that the effective coupling constant is $u\Lambda^{d-3}$ and therefore the upper critical dimensionality is $d_c=3$, i.e., the same as in equilibrium
at $T=0$. In the opposite case of a deep quench $\Omega_0 \gg \Lambda$, the effective coupling is
$\Omega_0 u \Lambda^{d-4}$ and, correspondingly, $d_c=4$ (see Appendix~\ref{app:app1}). This kind of \emph{dimensional crossover} is similar to the one occurring in equilibrium quantum systems upon varying $T$ \cite{Sachdev11,Sondhi97,Fisher88} (or in classical statistical systems in spatial confinement, see, e.g., Ref.~\cite{FiniteSize}).
{Equations \reff{eq:RG12} with constant $\Omega_0$, $d_c=4$ (i.e., for a deep quench), and $d<d_c$ admit a non-trivial, stable fixed point $Q_{\rm dy}(\Omega_0)\equiv (r^*_{\rm dy}(\Omega_0),u^*_{\rm dy}(\Omega_0))$ in the $(r,u)$-plane, which describes  a \emph{dynamical phase transition}. 
In particular, depending on the initial values $(r,u)$ of the parameters, after the non-universal transient of duration $t\simeq \Lambda^{-1}$ depicted in Fig.~\ref{fig:sketch}, their post-quench effective values $(r(\ell_t),u(\ell_t))$ determined by solving Eqs.~\reff{eq:RG12} may approach the fixed point $Q_{\rm dy}$ characterized by scaling behavior and aging. When $t$ exceeds $t^*$, $Q_{\rm dy}$ is generically destabilized as discussed further below.}
The RG Eqs.~\reff{eq:RG12}
are also very similar to those of this same quantum system in equilibrium at temperature $T$ (see, e.g., Ref.~\cite{Sachdev11}) --- with $\Omega_0$ playing the role of $T$ 
--- characterized by an \emph{equilibrium} fixed point $Q_{\rm eq}(T) \equiv (r^*_{\rm eq}(T),u^*_{\rm eq}(T))$.
Remarkably, up to this order in perturbation theory, the critical exponents $\nu$ derived 
by linearizing these two sets of RG equations around $Q_{\rm dy}$ and $Q_{\rm eq}$
are the same and equal $\nu_\text{eq} = 1/2 + \epsilon(N+2)/[4(N+8)] + {\mathcal O}(\epsilon^2)$, 
where $\epsilon \equiv d_c - d$ indicates the deviation from the upper critical dimensionality of the model.
One can actually define an \emph{effective temperature} $T = T_{\rm eff}(\Omega_0)$
such that the systems which are critical under equilibrium conditions are also critical after the quench.
This implies that the (linearized) critical lines of $Q_{\rm dy}$ and $Q_{\rm eq}$ in the $(r,u)$-plane are the same, though $Q_{\rm dy}(\Omega_0) \neq Q_{\rm eq}(T_{\rm eff}(\Omega_0))$. Only for $\Omega_0\gg\Lambda$, these two fixed points coincide, with $T_{\rm eff} = \Omega_0/4$ 
and  $r^*_{\rm dy}(\Omega_0) = r^*_{\rm eq}(T_{\rm eff}) = - \epsilon \Lambda^2 (N+2)/[2(N+8)] + {\mathcal O}(\epsilon^2)$.
In passing, we mention that the same happens also for $\Omega_0\ll \Lambda$.
In this respect and up to this order in perturbation theory, the dynamical transition (in the notion of Refs.~\cite{Sciolla2010,Sciolla11,Gambassi2011}) has some of the features of the equilibrium transition occurring at $T_{\rm eff}$, though differences could emerge at higher orders in perturbation theory or in quantities which depend on $Q_{\rm dy}$ or on the post-quench
distribution at short length scales, which is definitely not thermal \cite{Smacchia} (see further below).
It also remains to be seen whether  the $T_{\rm eff}$ defined above has any thermodynamic or dynamic role in the system, e.g., entering into fluctuation-dissipation relations~\cite{PhysRevB.84.212404,Foini11}.

The RG Eqs.~\reff{eq:RG12} have been derived under the assumption that inelastic scattering does not occur, at least
in the early stages of the evolution, and that the dynamical exponent keeps its initial value $z=1$.
In fact, up to this order in perturbation theory, the tadpole is the only relevant diagram which is responsible for the occurrence of elastic dephasing during the time evolution and, for a deep quench, it results in
the fixed point $Q_{\rm dy}$ discussed above.
However, the RG transformations also generate relevant dissipative terms  which are expected to drive the system to thermal equilibrium
\cite{Mitra11,Mitra12a}. In the present case, they appear as secular terms growing in time (see Appendix~\ref{app:app1}), eventually spoiling the perturbative expansion
(unless they are properly resummed \cite{Berges08,Tavora13,Lux13}),
and changing the dynamical exponent $z$ towards the diffusive value $z\simeq 2$.
Nonetheless, these terms, which are absent immediately after the quench and are therefore generated perturbatively, turn out to be small at short times $\Lambda t \lesssim \Lambda t^* = 1/(\Omega_0 u^*_{\rm dy}) \simeq \epsilon^{-1}$,
which include the range of times within which the short-time scaling behavior associated with $Q_{\rm dy}$ sets in (see Appendix~\ref{app:app1}).
Note that no dissipative terms are actually generated in the cases studied in Refs.~\cite{Sondhi2013, Sciolla2013, Smacchia}, namely in the $N\to \infty$ limit, because the relevant fluctuations are Gaussian. Accordingly,  the prethermal state is stable at all times and no thermalization occurs.

\section{Short-time scaling of various quantities} 
The emergence of a short-time scaling 
after a deep quench is clearly revealed by a perturbative calculation of $iG_K$ and $G_R$ for $k=0$, at the critical point $Q_{\rm dy}$. In fact, it turns out that for $t_2\ll t_1$ and up to ${\cal O}({u^*_{\rm dy}}^{\!\!\!2})$, $G_R(0,t_1\gg t_2) = - t_1 [1-\theta \ln(t_1/t_2)]
\simeq - t_1 (t_2/t_1)^\theta$ and, analogously, $iG_K(0,t_1,t_2) \simeq (\Omega_0/\Lambda^2)(\Lambda t_2)^{2-2\theta} (t_2/t_1)^{\theta-1}$, where 
\begin{equation}
\label{eq:theta}
\theta =  \frac{N+2}{N+8} \, \frac{\epsilon}{4} + {\cal O}(\epsilon^2).
\end{equation}
These algebraic dependences on time are similar but not identical to the ones observed in classical~\cite{Gambassi05} and quantum~\cite{Gagel14} systems undergoing \emph{aging} in contact with a thermal bath, with an \emph{initial-slip} exponent $\theta$.
{As in classical dissipative systems, 
$\theta$ emerges because the fields at $t=0$ acquire a different scaling dimension compared to those at
$t>0$, due to the breaking of TTI caused by the quench~\cite{Chiocchetta14}.
In the limit $N\to\infty$, Eq.~\reff{eq:theta} predicts the value $\theta_\infty = \epsilon/4 + \mathcal{O}(\epsilon^2)$ for the exponent $\theta$ 
of the very same model studied in Refs.~\cite{Sondhi2013, Sciolla2013, Smacchia}, although this universal short-time regime was overlooked by past studies, and constitutes a
central result of our paper.} 
The algebraic behavior of $G_{R,K}$ discussed above also appears
in the response function $-G_R$ as a function of the spatial distance $x=|\xx_1-\xx_2|$. For $u=0$, its expression $G_R^{(0)}(x,t_1-t_2)$ is TTI and shows typical light-cone dynamics by being enhanced at $x=t_1-t_2$ where $G_R^{(0)} \propto -\Lambda^3 \left[\Lambda (t_1-t_2)\right]^{-3/2}$ in $d=4$,
while decaying rapidly inside the light-cone for $x\ll t_1-t_2$, and being vanishingly small
outside it for $x\gg t_1-t_2$.
At one loop, $G_R$ is found to acquire an algebraic behavior for $t_2\ll t_1$, i.e., $G_R(x = {t_1} - {t_2},{t_2} \ll {t_1}) \simeq t_2^\theta \; t_1^{ - 3/2 + \epsilon/2}.$~\cite{Chiocchetta14}.
Analogously, the dynamics of the order parameter 
$\bar \vecphi(t)$ can be studied by adding a small symmetry-breaking field in the pre-quench Hamiltonian 
$H(\Omega_0^2,0)\rightarrow H(\Omega_0^2,0)-\int \dd^d x\, h_1\phi_1(x)$, which gives 
$\bar \phi_1(0^-) \equiv \phi_0 = h_1/\Omega_0^2\ll 1$.
The time evolution of $\bar \phi_1$ due to the post-quench Hamiltonian $H$ in Eq.~\reff{eq:Hamiltonian} (with no symmetry-breaking field) is determined by $[\partial_t^2 + M^2(t) - u{\bar \phi_1}^2/(3N)]\bar\phi_1(t)=0$
where $M^2(t) \simeq r + u\bar \phi^2_1/(2N) + u (N+2)iG_K(x=0,t,t)/(12N)$.  
At criticality $r=r_{\rm dy}^*(\Omega_0)$ and for times 
such that $\Lambda^{-1}\ll t\ll t_i$ where $\Lambda t_i\sim {\cal O}(|\phi_0|^{-1})$ one finds
$M^2(t)\simeq \theta/t^2$ and therefore $\bar\phi_1\simeq \phi_0 t^{\theta}$~\cite{Chiocchetta14}, i.e., the short-time evolution of  $\bar\phi_1$ is controlled by $\theta$ and corresponds to an initial \emph{increase} of the order with time.
{If the quench occurs slightly away from criticality, with $r = \delta r + r^*_{\rm dy}$, the short-time algebraic laws discussed above turn out to be modulated by oscillations of period $\propto |\delta r|^{-\nu z}$ \cite{Chiocchetta14}.}
%
%
\begin{figure}
\begin{center}
\includegraphics[width=0.32\textwidth]{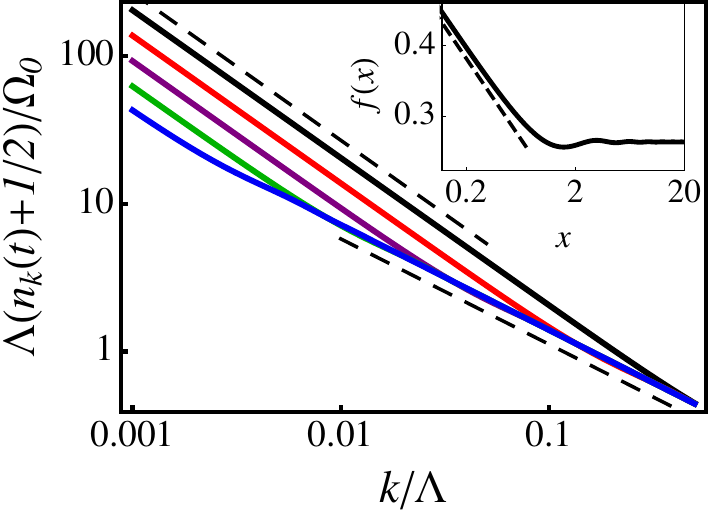}
\caption{(Color online) Momentum distribution $n_k$ after the quench, as a function of $k/\Lambda \ll 1$ for
$\Lambda t = 2$, 8, 32, 128, 512 (solid lines, from top to bottom). The algebraic short- and long-time behaviors of $n_k$ are highlighted by the upper $\sim k^{-1}$  and lower $\sim k^{-1+2\theta}$ dashed lines, respectively. The inset shows a log-log plot of the scaling function $f(x)$, which approaches $\sim x^{-2\theta}$ for $x\lesssim 1$ (dashed line). 
With the purpose of highlighting the crossover, we set $\epsilon=2$ in the perturbative expressions of these curves.}
\label{fig:nk}
\end{center}
\end{figure}
%
%

Remarkably, the momentum distribution $n_k$ of the quasi-particles also shows signatures of the exponent $\theta$ in the dependence on $k$ at criticality.
Immediately after the quench, $n_k$ takes the expected form of a GGE with a momentum-dependent effective temperature $T_{\rm eff}^k$ \cite{Calabrese06,Calabrese07}
which becomes independent of $k$ and equal to $T_{\rm eff}(\Omega_0)$ for deep quenches. Interactions eventually modify this behavior. In particular,
for a deep quench at the critical point $Q_{\rm dy}$, 
a perturbative calculation yields
$n_k(t) + 1/2 = (\Omega_0/\Lambda) (\Lambda/k)^{1-2\theta} f(kt)$, where the scaling function $f$ can be consistently estimated up to ${\cal O}(\epsilon)$
as the exponential of the one-loop correction and is such that $f(x\ll 1) \simeq x^{- 2 \theta}$, with a finite value for $x\gg 1$.
Accordingly, for fixed $t$,
$n_k(t) + 1/2$ as a function of $k$ crosses over from an algebraic behavior $\sim k^{-1} t^{-2\theta}$ for $k\lesssim t^{-1}$ to $\sim k^{-1+2\theta}$ for $k\gtrsim t^{-1}$.  This crossover is shown in Fig.~\ref{fig:nk} along with a plot of $f(x)$.
It is interesting to note that the dynamics of $n_k(t)$ in Fig.~\ref{fig:nk} closely resemble
the one observed at non-thermal fixed points (see, e.g., Ref.~\cite{PhysRevA.88.063615}).

{The scaling properties of $G_{R,K}$ discussed above bear remarkable differences compared to those in the classical case: for example, $G_R$ decreases $\propto t_2^\theta$ upon decreasing the smaller time $t_2$, whereas the opposite happens in the corresponding classical response function \cite{Gambassi05}. 
Nonetheless, the algebraic time dependence of $\bar \vecphi$ 
is the same as in the classical case and, in addition, the corresponding  exponent $\theta$ 
has the same value up to one-loop in spite of the fact that the dynamics are significantly different.
Indeed here the dynamical exponent is $z=1$ and energy is conserved, whereas $z>1$ in the classical case with a thermal bath.} 

The universal short-time behavior described here could be investigated, for the $O(N=2)$ universality class, in experimental realizations of the Bose-Hubbard model via ultra-cold atoms in optical lattices \cite{GreinerMandel02,Bakr2009,Bakr2010}. Alternatively,  the relative phase of tunnel-coupled condensates is known to be effectively described by Eq.~\reff{eq:Hamiltonian} with $N=1$ and in $d=1$ its dynamics 
has already been successfully studied in experiments \cite{Betz2011, Langen2013}. Similar protocols can also be adapted for fluids of light in non-linear optical systems \cite{Larre2014}. Finally, recent experimental realizations of systems with 
$SU(N)$ symmetry~\cite{Gorshkov2010, Zhang14} could be used in order to investigate the emergence of a short-time universal collective behavior in systems governed by an effective theory different from Eq.~\reff{eq:Hamiltonian}.

\section{Conclusions}
The RG analysis presented here demonstrates in a simple setting the emergence of a novel scaling behavior after a deep quench of an isolated quantum system. 
This phenomenon, due entirely to elastic dephasing, is an example of a macroscopic short-time
non-thermal fixed point; the corresponding behavior of various physical observables is controlled by a
universal exponent $\theta$, which we calculated at the first order in a dimensional expansion [see Eq.~\reff{eq:theta}].
The non-thermal fixed point is eventually destabilized towards a thermal regime, driven by dissipative terms generated in the effective action.

As the scaling regime unveiled here occurs at macroscopic short times, its numerical investigation should not be hampered by the computational limitations which typically prevent the investigation of the post-quench dynamics at long times.

\acknowledgments
The authors thank I.~Carusotto, M.~Marcuzzi, and A.~Silva for invaluable discussions. MT and AM were supported by
NSF-DMR 1303177.

\noindent\emph{Note.-- }A. C. and M. T. contributed equally to this work.

\appendix

\section{Momentum-shell renormalization group for the quench} 
\label{app:app1}
In this Appendix, we report the details of the RG analysis of a quench to the dynamical critical point of the model. For the sake of simplicity, we will consider the case with $N=1$, from which the generalization to a generic $N$ follows straightforwardly. In order to develop our RG analysis, it is convenient to introduce the functional formulation of the Keldysh formalism~\cite{Kamenevbook}, where any expectation value can be calculated as:
\begin{equation}
\label{eq:KeldyshAction}
\langle \mathcal{O}(t)\rangle = \int \mathcal{D}\phi \;\mathcal{O}[\phi_f(t)]\, \ee^{iS_K},
\end{equation}
where $S_K=S_K[\phi_f,\phi_b]$ is the so-called Keldysh action, while $\mathcal{D}\phi \equiv D\phi_f D\phi_b]$ is the functional measure. The two fields involved, denoted as \emph{forward} ($\phi_f$) and \emph{backward} ($\phi_b$), corresponds to the degrees of freedom defined, respectively, on the forward and backward branch of the Schwinger-Keldysh contour.
In the following it will be convenient to work with a linear combination of them, namely the \emph{classical} ($\phi_c$) and \emph{quantum} ($\phi_q$) fields, defined as~\cite{Kamenevbook} $\phi_c = (\phi_f+\phi_b)/\sqrt{2}$ and $\phi_q = (\phi_f-\phi_b)/\sqrt{2}$: in fact, in this basis, the retarded and Keldysh Green's functions (cf. Eqs. (2) and (3)) acquire a particularly simple form: 
\begin{align}
iG_R(|\xx-\xx'|,t,t') & =  \langle \phi_c(\xx,t)\phi_q(\xx',t') \rangle, \\
iG_K(|\xx-\xx'|,t,t') &  =  \langle \phi_c(\xx,t)\phi_c(\xx',t') \rangle.
\end{align}
For the Hamiltonian considered in Eq. (1), $S_K$ reads:
\begin{align}
\label{eq:SK}
S_K   & = S_\text{initial} +  \int_\xx \,\int_0^{+\infty}\dd t\,\left[\dot{\phi}_q\dot{\phi}_c -  (\nabla \phi_q)(\nabla \phi_c) \right.\nonumber \\
		& \left. \qquad \qquad   - r\,\phi_q\phi_c  - \frac{u_c}{4!}2\phi_q\phi_c^3 - \frac{u_q}{4!} 2\phi_q^3\phi_c\right],
\end{align}
where $\int_\xx \equiv \int \dd^d x$. Here $S_\text{initial}$ encodes the information about the initial state, while the remaining part is related to the post-quench Hamiltonian which rules the dynamics of the system for $t>0$; the explicit form of $S_\text{initial}$ is not needed in the following discussion and will be reported elsewhere~\cite{Chiocchetta14}. Note that the interaction term $\propto u\phi^4$ in Eq. \eqref{eq:Hamiltonian} is here represented by two terms, denoted as the \emph{classical} ($u_c$) and \emph{quantum} ($u_q$) vertices. While in principle $u_c = u_q = u$, we allow them to be different in order to discuss their RG flow.     

In the next sections we will derive the renormalization-group equations for the couplings in $S_K$. These equations will be derived by using a momentum-shell integration very similar to the one discussed in Refs.~\cite{Mitra11,Mitra12a}, which is based on Wilson's RG~\cite{Wilson74}. The calculations amount to a perturbation theory around the Gaussian point $u_c = u_q = 0$, justified in view of the eventual dimensional expansion around the upper critical dimension.

\subsection{RG equations}
In order to make finite any quantity computed from Eq. \eqref{eq:SK}, it is necessary to regularize the theory by curing the ultra-violet divergences of the integrals involved. More precisely, we will consider a sharp cut-off by requiring the Fourier components of the fields $\phi_{c,q}(\kk) = \int_\xx \phi_{c,q}(\xx)\ee^{-i\kk\cdot\xx}$ to vanish for momenta $k = |\vec{k}| > \Lambda$, where $\Lambda$ is a momentum scale which is related to the inverse of the smallest length-scale of an underlying microscopic model.

To perform the RG transformation, each field $\phi$ is decomposed in slow and fast components as $\phi = \phi_> + \phi_<$, where the slow component $\phi_<$ involves modes within the range $0\leq k < \Lambda -d\Lambda$, while the fast one $\phi_>$ involves modes within the momentum shell $\Lambda -d\Lambda \leq k \leq \Lambda$, where $d\Lambda \ll \Lambda$ is the thickness of the shell. Then, the fast modes are integrated out and consequently new terms are generated in the effective action for the slow modes. The integration of fast modes can be done by expanding the exponential weight to second order in the interaction terms, averaging fast modes over their Gaussian action $S_\text{Gauss}$ and finally re-exponentiating via a cumulant expansion:          
\begin{align}
\int \mathcal{D} &\phi_> \mathcal{D}\phi_< \;  \ee^{i( S_\mathrm{Gauss} +  S_\mathrm{int})} \nonumber \\
 & \simeq \int \mathcal{D} \phi_> \mathcal{D}\phi_<\; \ee^{i S_\mathrm{Gauss}} \left( 1+ i\, S_\mathrm{int} -\frac{1}{2} S_\mathrm{int}^2 \right) \nonumber\\ 
												& = \int \mathcal{D}\phi_<\; \ee^{iS^<_\text{Gauss}}\left[1 + i\, \langle S_{\mathrm{int}}\rangle_>  -\frac{1}{2}\langle S_{\mathrm{int}}^2\rangle_>\right] \nonumber\\
												& \simeq \int \mathcal{D}\phi_<\; \ee^{iS^<_\text{Gauss} + i\, \langle S_{\mathrm{int}}\rangle_>  -\frac{1}{2}\langle S_{\mathrm{int}}^2\rangle_>^c  },
\end{align}
where $ S_{\mathrm{int}} = -2u_c\int \phi^3_c\phi_q/4! - 2u_q\int \phi^3_q\phi_c/4!$ for the action in Eq. \eqref{eq:SK}. $\langle\dots\rangle_>$ denotes the expectation value with respect to the Gaussian action of the fast fields, while the superscript $c$ specifies that only connected terms are considered. The actual calculation can be easily performed by using the Wick's theorem to decompose every higher-order correlation function into products of the Gaussian Green's functions (see Eqs. (2) and (3)). After the integration, in order to restore the initial value of the cut-off $\Lambda$, coordinates and fields are rescaled as $x \to b x$, $t \to b^z t$ and  $\phi_{c,q} \to b^{\zeta_{c,q}}\phi_{c,q} $, where $b = \Lambda/(\Lambda - d\Lambda) \simeq 1 + d\Lambda/\Lambda$. Here $z$ is the dynamical critical exponent, while $\zeta_{c,q}$ correspond to the scaling dimension of the fields.  
Finally, the couplings $r'$ and $u'_{c,q}$ of the new action, which include contributions from both the integration of fast modes and the rescaling, are expressed in terms of the old couplings $r$ and $u_{c,q}$ in the form of recursion relations:
\begin{align}
\label{eq:RGrecursion1}
r' &= b^2\,\left(r + u_c I_1\right), \\
u_c' &= b^{2-2\zeta_c} \,u_c\, \left(1- u_c I_2\right), \label{eq:RGrecursion2}\\
u_q' &= b^{2-2\zeta_q} \,u_q\, \left(1- u_q I_2\right), \label{eq:RGrecursion3}
\end{align}
where we have set $z=1$, by requiring the coefficient of time- and spatial-derivatives in $S_K$ to be invariant under renormalization. 
The integrals $I_{1,2}$, which result from diagrams up to one-loop, depend on time $t$, because $G_K$ is not time-translational invariant. 
It is therefore convenient to decompose them in the time-independent and time-dependent parts as $I_{1,2}(t) = J^\infty_{1,2} + J_{1,2}(t)$, where 
\begin{align}
J^\infty_1  &= \frac{d\Lambda}{\Lambda} \frac{a_d}{8}\Lambda^d \frac{2\Lambda^2 + \Omega_0^2 + r}{(\Lambda^2 + r)\sqrt{\Lambda^2 + \Omega_0^2}},  \label{eq:integral_TI_1}\\
 J^\infty_2  &= \frac{d\Lambda}{\Lambda}  \frac{3}{8} a_d \Lambda^d \frac{\sqrt{\Lambda^2 +\Omega_0^2}}{(\Lambda^2 + r)^2} , \label{eq:integral_TI_2}
\end{align}
and
\begin{align}
J_1(t) & = \frac{d\Lambda}{\Lambda} \frac{a_d}{8} \Lambda^d \frac{r-\Omega_0^2}{(\Lambda^2 + r)\sqrt{\Lambda^2+\omega_0^2}} \cos(2t\sqrt{\Lambda^2 + r}), \label{eq:integral1_TD}\\
J_2(t) & = \frac{d\Lambda}{\Lambda}  \frac{3}{8} a_d \Lambda^d \frac{\sqrt{\Lambda^2+\Omega^2_0}}{(\Lambda^2+r)^2} \, \left[ - \cos\left(2t\sqrt{\Lambda^2+r} \right) \right. \nonumber\\ 
        & \left. \qquad +\frac{(r-\Omega_0^2)\sqrt{\Lambda^2+r}}{\Lambda^2 + \Omega_0^2}\, t\, \sin\left(2t\sqrt{\Lambda^2+\Omega_0^2} \right)    \right].\label{eq:integral2_TD}
\end{align}
$J_{1,2}(t)$ are fast oscillating functions of time which, in principle, contribute to the renormalization of the couplings. However, these oscillations should be regarded as an artifact of imposing a sharp cut-off in the momentum integrals involved in the calculation: if a smooth cut-off function is considered, instead they are replaced by functions which vanish smoothly upon increasing $\Lambda t$. Accordingly, for $\Lambda t \gg 1$, the contribution of $J_{1,2}(t)$ to the renormalization of the couplings become negligible, and after a time $t \simeq \Lambda^{-1}$ the recursion relations \eqref{eq:RGrecursion1}-\eqref{eq:RGrecursion3} become time-independent.              

In order to rewrite the RG recursion equations in a differential form, we introduce the infinitesimal dimensionless parameter 
$\delta \ell = d\Lambda/\Lambda$ and we retain only the first order in $\delta \ell$. 
Accordingly, from Eqs. \eqref{eq:RGrecursion1}, \eqref{eq:RGrecursion2} and \eqref{eq:RGrecursion3} we find the differential equations: 
 \begin{align}
\frac{d r}{d \ell} & = 2 r + u_c \frac{a_d}{8}\Lambda^d \frac{2\Lambda^2 + \Omega_0^2 + r}{(\Lambda^2 + r)\sqrt{\Lambda^2 + \Omega_0^2}}, \label{eq:RG_r}\\
\frac{d u_c}{d \ell} & = u_c\left[(d_c-d) -  u_c \,\frac{3}{8}a_d    \frac{\sqrt{\Lambda^2 +\Omega_0^2}}{(\Lambda^2 + r)^2} \right], \label{eq:RG_uc}\\
\frac{d u_q}{d \ell} & = u_q\left[(d_q-d) -  u_c \,\frac{3}{8}a_d    \frac{\sqrt{\Lambda^2 +\Omega_0^2}}{(\Lambda^2 + r)^2} \right], \label{eq:RG_uq}
 \end{align}
where we defined the classical and quantum upper critical dimensions $d_{c,q}$ as, respectively, $d_{c,q} = 2+d - 2\zeta_{c,q}$. 
The upper critical dimensions are thus determined from the (Gaussian) scaling dimensions $\zeta_{c,q}$ of the fields $\phi_{c,q}$, which in turn should be determined 
by requiring each term in $S_K$ to be dimensionless. However, as it is apparent from Eq. \eqref{eq:KeldyshAction}, 
the fields $\phi_{c,q}$ appear always in the combination $\phi_c\phi_q$, and therefore it is impossible to determine their scaling dimensions separately; 
nevertheless, $\zeta_c$ can be deduced from a direct inspection of the Gaussian correlation function $\langle \phi_c(\xx,t)\phi_c(\xx',t)\rangle = iG_K$. 
To this end, we focus on the case of a deep quench, where, as discussed in the main text, $G_K$ resembles the equilibrium one with an effective 
temperature $T_\text{eff} = \Omega_0/4$, namely
\begin{equation}
\label{eq:GK-deepquench}
\langle \phi_c(\xx,t)\phi_c(\xx',t)\rangle \sim \Omega_0 \int \dd^d k\,  \frac{\ee^{i\kk\cdot(\xx-\xx')}}{k^2}. 
\end{equation}
Since $\Omega_0$ plays the role of an effective temperature, we argue that it does not flow under RG transformations, in analogy to what happens 
for the temperature in equilibrium quantum systems \cite{Fisher1988}. Accordingly, from a simple rescaling in Eq. \eqref{eq:GK-deepquench}, 
we find $\zeta_c = (d-2)/2$ and, correspondingly, $\zeta_q = d/2$; when plugged into Eqs. \eqref{eq:RG_r}-\eqref{eq:RG_uq}, these values give
 \begin{align}
\frac{d r}{d \ell} & = 2 r + u_c \frac{a_d}{8}\Lambda^d \frac{2\Lambda^2 + \Omega_0^2 + r}{(\Lambda^2 + r)\sqrt{\Lambda^2 + \Omega_0^2}}, \label{eq:RG_r-b}\\
\frac{d u_c}{d \ell} & = u_c\left[(4-d) -  u_c \,\frac{3}{8}a_d    \frac{\sqrt{\Lambda^2 +\Omega_0^2}}{(\Lambda^2 + r)^2} \right], \label{eq:RG_uc-b}\\
\frac{d u_q}{d \ell} & = u_q\left[(2-d) -  u_c \,\frac{3}{8}a_d    \frac{\sqrt{\Lambda^2 +\Omega_0^2}}{(\Lambda^2 + r)^2} \right]. \label{eq:RG_uq-b}
 \end{align}
These equations show that the coupling $u_q$ of quantum vertex becomes irrelevant for $d>2$, while the coupling $u_c$of the classical vertex 
becomes irrelevant for $d>4$, thus fixing the upper critical dimension $d_c$ to $d_c=4$. Accordingly, for deep quenches, the upper critical dimension is the 
same as in the corresponding equilibrium case at finite temperature~\cite{Fisher1988,Sondhi97,Sachdev11}. 
We emphasize that, for $\Omega_0=0$, the Gaussian Keldysh Green's function reads         
\begin{equation}
\label{eq:GK-shallowquench}
\langle \phi_c(\xx,t)\phi_c(\xx',t)\rangle = \int \dd^d k\,  \frac{\ee^{i\kk\cdot(\xx-\xx')}}{k},
\end{equation}
from which one infers $\zeta_c =\zeta_q= (d-1)/2$. In this case, the quantum and classical vertices $u_c$ and $u_q$ have the same upper critical dimension which is $d_c=3$, similarly to the corresponding equilibrium system at $T=0$. This shows that $\Omega_0$ plays, for the dynamical phase transition, the same role that the temperature plays for the corresponding equilibrium quantum phase transition.   

\subsection{Dissipative and secular terms}
\label{sec:dissipative-terms}
The emergence of any dissipative mechanism is primarily related to the generation of terms like 
$g_2\phi_q^2$ and $g_4(\phi_q\phi_c)^2$ in $S_K$. In fact, they can be shown to correspond~\cite{Kamenevbook} to an effective external noise driving the dynamics of the relevant fields. On the basis of power counting in the case of deep quenches, the couplings $g_2$ and $g_4$ scale as
\begin{equation}
g_2 \sim \mu, \qquad g_4 \sim \mu^{3-d},
\end{equation}
where $\mu$ is an arbitrary momentum scale; accordingly $g_2$ is relevant for all spatial dimensions while $g_4$ becomes negligible for $d>3$. As a result, even if the dynamics starts from an action without these terms, they can be generated under renormalization and, in this case, their effect will dramatically change the properties of the theory, by inducing a crossover and a change of the scaling dimensions. Nevertheless, the renormalization of $g_2$ requires at least a two-loop correction which can be consequently neglected in our one-loop analysis. On the other hand, a correction to $g_2$ is actually generated at one loop, as 
\begin{align}
\label{eq:secular-term}
\delta g_4 & =  -i  \frac{a_d}{16}  u_c\, t\, \frac{\Lambda^d}{\Lambda^2 + r} \times \nonumber\\
& \times \left[u_q  - \frac{u_c}{2}\left( \frac{\Lambda^2 + r}{\Lambda^2 + \Omega_0^2} + \frac{\Lambda^2 + \Omega_0^2}{\Lambda^2 + r}   \right)\right] d \ell,
\end{align}
where oscillating terms have been neglected (see the discussion above). This correction increases upon increasing the time $t$ elapsed from the quench and therefore, 
even if it is irrelevant for $d>3$, it becomes eventually important, spoiling the perturbative expansion. This kind of linear growth is nothing but a secular term related 
to the simple perturbative approach. Although several techniques have been proposed in order to avoid this problem~\cite{Berges2004,Moeckel2008}, 
we emphasize that it dramatically affects only the long time properties of the system, rather than the short time ones that we are considering here. We can provide an heuristic 
estimate of the time $t^*$ at which such a term becomes relevant, by considering when $\delta g_4$ becomes of order $\mathcal{O}(\epsilon)$
at the critical point $u_c=u^*$ discussed in the main text. In this case $\delta g_4$ turns out to be is negligible for times $\Lambda t^* \lesssim \epsilon^{-1}$: 
accordingly, close to the upper critical dimension $\epsilon \to 0$, the dissipative vertex can be safely neglected at short times.


\bibliography{quenchnew2}

\begin{thebibliography}{90}%
\makeatletter
\providecommand \@ifxundefined [1]{%
 \@ifx{#1\undefined}
}%
\providecommand \@ifnum [1]{%
 \ifnum #1\expandafter \@firstoftwo
 \else \expandafter \@secondoftwo
 \fi
}%
\providecommand \@ifx [1]{%
 \ifx #1\expandafter \@firstoftwo
 \else \expandafter \@secondoftwo
 \fi
}%
\providecommand \natexlab [1]{#1}%
\providecommand \enquote  [1]{``#1''}%
\providecommand \bibnamefont  [1]{#1}%
\providecommand \bibfnamefont [1]{#1}%
\providecommand \citenamefont [1]{#1}%
\providecommand \href@noop [0]{\@secondoftwo}%
\providecommand \href [0]{\begingroup \@sanitize@url \@href}%
\providecommand \@href[1]{\@@startlink{#1}\@@href}%
\providecommand \@@href[1]{\endgroup#1\@@endlink}%
\providecommand \@sanitize@url [0]{\catcode `\\12\catcode `\$12\catcode
  `\&12\catcode `\#12\catcode `\^12\catcode `\_12\catcode `\%12\relax}%
\providecommand \@@startlink[1]{}%
\providecommand \@@endlink[0]{}%
\providecommand \url  [0]{\begingroup\@sanitize@url \@url }%
\providecommand \@url [1]{\endgroup\@href {#1}{\urlprefix }}%
\providecommand \urlprefix  [0]{URL }%
\providecommand \Eprint [0]{\href }%
\providecommand \doibase [0]{http://dx.doi.org/}%
\providecommand \selectlanguage [0]{\@gobble}%
\providecommand \bibinfo  [0]{\@secondoftwo}%
\providecommand \bibfield  [0]{\@secondoftwo}%
\providecommand \translation [1]{[#1]}%
\providecommand \BibitemOpen [0]{}%
\providecommand \bibitemStop [0]{}%
\providecommand \bibitemNoStop [0]{.\EOS\space}%
\providecommand \EOS [0]{\spacefactor3000\relax}%
\providecommand \BibitemShut  [1]{\csname bibitem#1\endcsname}%
\let\auto@bib@innerbib\@empty
\bibitem [{\citenamefont {Polkovnikov}\ \emph {et~al.}(2011)\citenamefont
  {Polkovnikov}, \citenamefont {Sengupta}, \citenamefont {Silva},\ and\
  \citenamefont {Vengalattore}}]{Polkovnikovrev}%
  \BibitemOpen
  \bibfield  {author} {\bibinfo {author} {\bibfnamefont {A.}~\bibnamefont
  {Polkovnikov}}, \bibinfo {author} {\bibfnamefont {K.}~\bibnamefont
  {Sengupta}}, \bibinfo {author} {\bibfnamefont {A.}~\bibnamefont {Silva}}, \
  and\ \bibinfo {author} {\bibfnamefont {M.}~\bibnamefont {Vengalattore}},\
  }\href {\doibase 10.1103/RevModPhys.83.863} {\bibfield  {journal} {\bibinfo
  {journal} {Rev. Mod. Phys.}\ }\textbf {\bibinfo {volume} {83}},\ \bibinfo
  {pages} {863} (\bibinfo {year} {2011})}\BibitemShut {NoStop}%
\bibitem [{\citenamefont {Lamacraft}\ and\ \citenamefont
  {Moore}(2012)}]{Lamacraft12}%
  \BibitemOpen
  \bibfield  {author} {\bibinfo {author} {\bibfnamefont {A.}~\bibnamefont
  {Lamacraft}}\ and\ \bibinfo {author} {\bibfnamefont {J.}~\bibnamefont
  {Moore}},\ }in\ \href@noop {} {\emph {\bibinfo {booktitle} {Ultracold Bosonic
  and Fermionic Gases}}},\ \bibinfo {editor} {edited by\ \bibinfo {editor}
  {\bibfnamefont {A.}~\bibnamefont {Fetter}}, \bibinfo {editor} {\bibfnamefont
  {K.}~\bibnamefont {Levin}}, \ and\ \bibinfo {editor} {\bibfnamefont
  {D.}~\bibnamefont {Stamper-Kurn}}}\ (\bibinfo  {publisher} {Elsevier,
  Amsterdam},\ \bibinfo {year} {2012})\ Chap.~\bibinfo {chapter}
  {7}\BibitemShut {NoStop}%
\bibitem [{\citenamefont {Yukalov}(2011)}]{Yukalov11}%
  \BibitemOpen
  \bibfield  {author} {\bibinfo {author} {\bibfnamefont {V.}~\bibnamefont
  {Yukalov}},\ }\href {\doibase 10.1002/lapl.201110002} {\bibfield  {journal}
  {\bibinfo  {journal} {Laser Phys. Lett.}\ }\textbf {\bibinfo {volume} {8}},\
  \bibinfo {pages} {485} (\bibinfo {year} {2011})}\BibitemShut {NoStop}%
\bibitem [{\citenamefont {Bloch}\ \emph {et~al.}(2008)\citenamefont {Bloch},
  \citenamefont {Dalibard},\ and\ \citenamefont {Zwerger}}]{Bloch08}%
  \BibitemOpen
  \bibfield  {author} {\bibinfo {author} {\bibfnamefont {I.}~\bibnamefont
  {Bloch}}, \bibinfo {author} {\bibfnamefont {J.}~\bibnamefont {Dalibard}}, \
  and\ \bibinfo {author} {\bibfnamefont {W.}~\bibnamefont {Zwerger}},\ }\href
  {\doibase 10.1103/RevModPhys.80.885} {\bibfield  {journal} {\bibinfo
  {journal} {Rev. Mod. Phys.}\ }\textbf {\bibinfo {volume} {80}},\ \bibinfo
  {pages} {885} (\bibinfo {year} {2008})}\BibitemShut {NoStop}%
\bibitem [{\citenamefont {Greiner}\ \emph
  {et~al.}(2002{\natexlab{a}})\citenamefont {Greiner}, \citenamefont {Mandel},
  \citenamefont {Esslinger}, \citenamefont {H\"ansch},\ and\ \citenamefont
  {Bloch}}]{GreinerMandel02}%
  \BibitemOpen
  \bibfield  {author} {\bibinfo {author} {\bibfnamefont {M.}~\bibnamefont
  {Greiner}}, \bibinfo {author} {\bibfnamefont {O.}~\bibnamefont {Mandel}},
  \bibinfo {author} {\bibfnamefont {T.}~\bibnamefont {Esslinger}}, \bibinfo
  {author} {\bibfnamefont {T.~W.}\ \bibnamefont {H\"ansch}}, \ and\ \bibinfo
  {author} {\bibfnamefont {I.}~\bibnamefont {Bloch}},\ }\href {\doibase
  10.1038/415039a} {\bibfield  {journal} {\bibinfo  {journal} {Nature}\
  }\textbf {\bibinfo {volume} {415}},\ \bibinfo {pages} {39} (\bibinfo {year}
  {2002}{\natexlab{a}})}\BibitemShut {NoStop}%
\bibitem [{\citenamefont {Greiner}\ \emph
  {et~al.}(2002{\natexlab{b}})\citenamefont {Greiner}, \citenamefont {Mandel},
  \citenamefont {Hansch},\ and\ \citenamefont {Bloch}}]{GMHB-2002}%
  \BibitemOpen
  \bibfield  {author} {\bibinfo {author} {\bibfnamefont {M.}~\bibnamefont
  {Greiner}}, \bibinfo {author} {\bibfnamefont {O.}~\bibnamefont {Mandel}},
  \bibinfo {author} {\bibfnamefont {T.~W.}\ \bibnamefont {Hansch}}, \ and\
  \bibinfo {author} {\bibfnamefont {I.}~\bibnamefont {Bloch}},\ }\href@noop {}
  {\bibfield  {journal} {\bibinfo  {journal} {Nature}\ }\textbf {\bibinfo
  {volume} {419}},\ \bibinfo {pages} {51} (\bibinfo {year}
  {2002}{\natexlab{b}})}\BibitemShut {NoStop}%
\bibitem [{\citenamefont {Deutsch}(1991)}]{Deutsch91}%
  \BibitemOpen
  \bibfield  {author} {\bibinfo {author} {\bibfnamefont {J.~M.}\ \bibnamefont
  {Deutsch}},\ }\href {\doibase 10.1103/PhysRevA.43.2046} {\bibfield  {journal}
  {\bibinfo  {journal} {Phys. Rev. A}\ }\textbf {\bibinfo {volume} {43}},\
  \bibinfo {pages} {2046} (\bibinfo {year} {1991})}\BibitemShut {NoStop}%
\bibitem [{\citenamefont {Srednicki}(1994)}]{Srednicki94}%
  \BibitemOpen
  \bibfield  {author} {\bibinfo {author} {\bibfnamefont {M.}~\bibnamefont
  {Srednicki}},\ }\href {\doibase 10.1103/PhysRevE.50.888} {\bibfield
  {journal} {\bibinfo  {journal} {Phys. Rev. E}\ }\textbf {\bibinfo {volume}
  {50}},\ \bibinfo {pages} {888} (\bibinfo {year} {1994})}\BibitemShut
  {NoStop}%
\bibitem [{\citenamefont {Rigol}\ \emph {et~al.}(2008)\citenamefont {Rigol},
  \citenamefont {Dunjko},\ and\ \citenamefont {Olshanii}}]{RigolDunjko08}%
  \BibitemOpen
  \bibfield  {author} {\bibinfo {author} {\bibfnamefont {M.}~\bibnamefont
  {Rigol}}, \bibinfo {author} {\bibfnamefont {V.}~\bibnamefont {Dunjko}}, \
  and\ \bibinfo {author} {\bibfnamefont {M.}~\bibnamefont {Olshanii}},\ }\href
  {\doibase 10.1038/nature06838} {\bibfield  {journal} {\bibinfo  {journal}
  {Nature}\ }\textbf {\bibinfo {volume} {452}},\ \bibinfo {pages} {854}
  (\bibinfo {year} {2008})}\BibitemShut {NoStop}%
\bibitem [{\citenamefont {Berges}\ \emph {et~al.}(2004)\citenamefont {Berges},
  \citenamefont {Bors\'anyi},\ and\ \citenamefont {Wetterich}}]{Berges04}%
  \BibitemOpen
  \bibfield  {author} {\bibinfo {author} {\bibfnamefont {J.}~\bibnamefont
  {Berges}}, \bibinfo {author} {\bibfnamefont {S.}~\bibnamefont {Bors\'anyi}},
  \ and\ \bibinfo {author} {\bibfnamefont {C.}~\bibnamefont {Wetterich}},\
  }\href {\doibase 10.1103/PhysRevLett.93.142002} {\bibfield  {journal}
  {\bibinfo  {journal} {Phys. Rev. Lett.}\ }\textbf {\bibinfo {volume} {93}},\
  \bibinfo {pages} {142002} (\bibinfo {year} {2004})}\BibitemShut {NoStop}%
\bibitem [{\citenamefont {Kitagawa}\ \emph {et~al.}(2011)\citenamefont
  {Kitagawa}, \citenamefont {Imambekov}, \citenamefont {Schmiedmayer},\ and\
  \citenamefont {Demler}}]{Kitagawa11a}%
  \BibitemOpen
  \bibfield  {author} {\bibinfo {author} {\bibfnamefont {T.}~\bibnamefont
  {Kitagawa}}, \bibinfo {author} {\bibfnamefont {A.}~\bibnamefont {Imambekov}},
  \bibinfo {author} {\bibfnamefont {J.}~\bibnamefont {Schmiedmayer}}, \ and\
  \bibinfo {author} {\bibfnamefont {E.}~\bibnamefont {Demler}},\ }\href@noop {}
  {\bibfield  {journal} {\bibinfo  {journal} {New J. Phys.}\ }\textbf {\bibinfo
  {volume} {13}},\ \bibinfo {pages} {073018} (\bibinfo {year}
  {2011})}\BibitemShut {NoStop}%
\bibitem [{\citenamefont {Gring}\ \emph {et~al.}(2012)\citenamefont {Gring},
  \citenamefont {Kuhnert}, \citenamefont {Langen}, \citenamefont {Kitagawa},
  \citenamefont {Rauer}, \citenamefont {Schreitl}, \citenamefont {Mazets},
  \citenamefont {Smith}, \citenamefont {Demler},\ and\ \citenamefont
  {Schmiedmayer}}]{Gring12}%
  \BibitemOpen
  \bibfield  {author} {\bibinfo {author} {\bibfnamefont {M.}~\bibnamefont
  {Gring}}, \bibinfo {author} {\bibfnamefont {M.}~\bibnamefont {Kuhnert}},
  \bibinfo {author} {\bibfnamefont {T.}~\bibnamefont {Langen}}, \bibinfo
  {author} {\bibfnamefont {T.}~\bibnamefont {Kitagawa}}, \bibinfo {author}
  {\bibfnamefont {B.}~\bibnamefont {Rauer}}, \bibinfo {author} {\bibfnamefont
  {M.}~\bibnamefont {Schreitl}}, \bibinfo {author} {\bibfnamefont
  {I.}~\bibnamefont {Mazets}}, \bibinfo {author} {\bibfnamefont {D.~A.}\
  \bibnamefont {Smith}}, \bibinfo {author} {\bibfnamefont {E.}~\bibnamefont
  {Demler}}, \ and\ \bibinfo {author} {\bibfnamefont {J.}~\bibnamefont
  {Schmiedmayer}},\ }\href@noop {} {\bibfield  {journal} {\bibinfo  {journal}
  {Science}\ }\textbf {\bibinfo {volume} {337}},\ \bibinfo {pages} {1318}
  (\bibinfo {year} {2012})}\BibitemShut {NoStop}%
\bibitem [{\citenamefont {Langen}\ \emph {et~al.}(2013)\citenamefont {Langen},
  \citenamefont {Geiger}, \citenamefont {Kuhnert}, \citenamefont {Rauer},\ and\
  \citenamefont {Schmiedmayer}}]{Langen2013}%
  \BibitemOpen
  \bibfield  {author} {\bibinfo {author} {\bibfnamefont {T.}~\bibnamefont
  {Langen}}, \bibinfo {author} {\bibfnamefont {R.}~\bibnamefont {Geiger}},
  \bibinfo {author} {\bibfnamefont {M.}~\bibnamefont {Kuhnert}}, \bibinfo
  {author} {\bibfnamefont {B.}~\bibnamefont {Rauer}}, \ and\ \bibinfo {author}
  {\bibfnamefont {J.}~\bibnamefont {Schmiedmayer}},\ }\href@noop {} {\bibfield
  {journal} {\bibinfo  {journal} {Nat. Phys.}\ }\textbf {\bibinfo {volume}
  {9}},\ \bibinfo {pages} {640} (\bibinfo {year} {2013})}\BibitemShut {NoStop}%
\bibitem [{\citenamefont {Kollar}\ \emph {et~al.}(2011)\citenamefont {Kollar},
  \citenamefont {Wolf},\ and\ \citenamefont {Eckstein}}]{Kollar11}%
  \BibitemOpen
  \bibfield  {author} {\bibinfo {author} {\bibfnamefont {M.}~\bibnamefont
  {Kollar}}, \bibinfo {author} {\bibfnamefont {F.~A.}\ \bibnamefont {Wolf}}, \
  and\ \bibinfo {author} {\bibfnamefont {M.}~\bibnamefont {Eckstein}},\ }\href
  {\doibase 10.1103/PhysRevB.84.054304} {\bibfield  {journal} {\bibinfo
  {journal} {Phys. Rev. B}\ }\textbf {\bibinfo {volume} {84}},\ \bibinfo
  {pages} {054304} (\bibinfo {year} {2011})}\BibitemShut {NoStop}%
\bibitem [{\citenamefont {Moeckel}\ and\ \citenamefont
  {Kehrein}(2008{\natexlab{a}})}]{Kehrein08}%
  \BibitemOpen
  \bibfield  {author} {\bibinfo {author} {\bibfnamefont {M.}~\bibnamefont
  {Moeckel}}\ and\ \bibinfo {author} {\bibfnamefont {S.}~\bibnamefont
  {Kehrein}},\ }\href {\doibase 10.1103/PhysRevLett.100.175702} {\bibfield
  {journal} {\bibinfo  {journal} {Phys. Rev. Lett.}\ }\textbf {\bibinfo
  {volume} {100}},\ \bibinfo {pages} {175702} (\bibinfo {year}
  {2008}{\natexlab{a}})}\BibitemShut {NoStop}%
\bibitem [{\citenamefont {Moeckel}\ and\ \citenamefont
  {Kehrein}(2009)}]{Moeckel09}%
  \BibitemOpen
  \bibfield  {author} {\bibinfo {author} {\bibfnamefont {M.}~\bibnamefont
  {Moeckel}}\ and\ \bibinfo {author} {\bibfnamefont {S.}~\bibnamefont
  {Kehrein}},\ }\href {\doibase http://dx.doi.org/10.1016/j.aop.2009.03.009}
  {\bibfield  {journal} {\bibinfo  {journal} {Annals of Physics}\ }\textbf
  {\bibinfo {volume} {324}},\ \bibinfo {pages} {2146 } (\bibinfo {year}
  {2009})}\BibitemShut {NoStop}%
\bibitem [{\citenamefont {Moeckel}\ and\ \citenamefont
  {Kehrein}(2010)}]{Moeckel10}%
  \BibitemOpen
  \bibfield  {author} {\bibinfo {author} {\bibfnamefont {M.}~\bibnamefont
  {Moeckel}}\ and\ \bibinfo {author} {\bibfnamefont {S.}~\bibnamefont
  {Kehrein}},\ }\href@noop {} {\bibfield  {journal} {\bibinfo  {journal} {New
  J. Phys.}\ }\textbf {\bibinfo {volume} {12}},\ \bibinfo {pages} {055016}
  (\bibinfo {year} {2010})}\BibitemShut {NoStop}%
\bibitem [{\citenamefont {Marino}\ and\ \citenamefont
  {Silva}(2012)}]{Marino12}%
  \BibitemOpen
  \bibfield  {author} {\bibinfo {author} {\bibfnamefont {J.}~\bibnamefont
  {Marino}}\ and\ \bibinfo {author} {\bibfnamefont {A.}~\bibnamefont {Silva}},\
  }\href {\doibase 10.1103/PhysRevB.86.060408} {\bibfield  {journal} {\bibinfo
  {journal} {Phys. Rev. B}\ }\textbf {\bibinfo {volume} {86}},\ \bibinfo
  {pages} {060408} (\bibinfo {year} {2012})}\BibitemShut {NoStop}%
\bibitem [{\citenamefont {Mitra}(2013)}]{Mitra13a}%
  \BibitemOpen
  \bibfield  {author} {\bibinfo {author} {\bibfnamefont {A.}~\bibnamefont
  {Mitra}},\ }\href {\doibase 10.1103/PhysRevB.87.205109} {\bibfield  {journal}
  {\bibinfo  {journal} {Phys. Rev. B}\ }\textbf {\bibinfo {volume} {87}},\
  \bibinfo {pages} {205109} (\bibinfo {year} {2013})}\BibitemShut {NoStop}%
\bibitem [{\citenamefont {van~den Worm}\ \emph {et~al.}(2013)\citenamefont
  {van~den Worm}, \citenamefont {Sawyer}, \citenamefont {Bollinger},\ and\
  \citenamefont {Kastner}}]{Worm13}%
  \BibitemOpen
  \bibfield  {author} {\bibinfo {author} {\bibfnamefont {M.}~\bibnamefont
  {van~den Worm}}, \bibinfo {author} {\bibfnamefont {B.~C.}\ \bibnamefont
  {Sawyer}}, \bibinfo {author} {\bibfnamefont {J.~J.}\ \bibnamefont
  {Bollinger}}, \ and\ \bibinfo {author} {\bibfnamefont {M.}~\bibnamefont
  {Kastner}},\ }\href@noop {} {\bibfield  {journal} {\bibinfo  {journal} {New
  J. Phys.}\ }\textbf {\bibinfo {volume} {15}},\ \bibinfo {pages} {083007}
  (\bibinfo {year} {2013})}\BibitemShut {NoStop}%
\bibitem [{\citenamefont {Marcuzzi}\ \emph {et~al.}(2013)\citenamefont
  {Marcuzzi}, \citenamefont {Marino}, \citenamefont {Gambassi},\ and\
  \citenamefont {Silva}}]{Marcuzzi13}%
  \BibitemOpen
  \bibfield  {author} {\bibinfo {author} {\bibfnamefont {M.}~\bibnamefont
  {Marcuzzi}}, \bibinfo {author} {\bibfnamefont {J.}~\bibnamefont {Marino}},
  \bibinfo {author} {\bibfnamefont {A.}~\bibnamefont {Gambassi}}, \ and\
  \bibinfo {author} {\bibfnamefont {A.}~\bibnamefont {Silva}},\ }\href
  {\doibase 10.1103/PhysRevLett.111.197203} {\bibfield  {journal} {\bibinfo
  {journal} {Phys. Rev. Lett.}\ }\textbf {\bibinfo {volume} {111}},\ \bibinfo
  {pages} {197203} (\bibinfo {year} {2013})}\BibitemShut {NoStop}%
\bibitem [{\citenamefont {Kinoshita}\ \emph {et~al.}(2006)\citenamefont
  {Kinoshita}, \citenamefont {Wenger},\ and\ \citenamefont {Weiss}}]{Weiss06}%
  \BibitemOpen
  \bibfield  {author} {\bibinfo {author} {\bibfnamefont {T.}~\bibnamefont
  {Kinoshita}}, \bibinfo {author} {\bibfnamefont {T.}~\bibnamefont {Wenger}}, \
  and\ \bibinfo {author} {\bibfnamefont {D.~S.}\ \bibnamefont {Weiss}},\
  }\href@noop {} {\bibfield  {journal} {\bibinfo  {journal} {Nature (London)}\
  }\textbf {\bibinfo {volume} {440}},\ \bibinfo {pages} {900} (\bibinfo {year}
  {2006})}\BibitemShut {NoStop}%
\bibitem [{\citenamefont {Rigol}\ \emph {et~al.}(2007)\citenamefont {Rigol},
  \citenamefont {Dunjko}, \citenamefont {Yurovsky},\ and\ \citenamefont
  {Olshanii}}]{Rigol07}%
  \BibitemOpen
  \bibfield  {author} {\bibinfo {author} {\bibfnamefont {M.}~\bibnamefont
  {Rigol}}, \bibinfo {author} {\bibfnamefont {V.}~\bibnamefont {Dunjko}},
  \bibinfo {author} {\bibfnamefont {V.}~\bibnamefont {Yurovsky}}, \ and\
  \bibinfo {author} {\bibfnamefont {M.}~\bibnamefont {Olshanii}},\ }\href
  {\doibase 10.1103/PhysRevLett.98.050405} {\bibfield  {journal} {\bibinfo
  {journal} {Phys. Rev. Lett.}\ }\textbf {\bibinfo {volume} {98}},\ \bibinfo
  {pages} {050405} (\bibinfo {year} {2007})}\BibitemShut {NoStop}%
\bibitem [{\citenamefont {Iucci}\ and\ \citenamefont
  {Cazalilla}(2009)}]{PhysRevA.80.063619}%
  \BibitemOpen
  \bibfield  {author} {\bibinfo {author} {\bibfnamefont {A.}~\bibnamefont
  {Iucci}}\ and\ \bibinfo {author} {\bibfnamefont {M.~A.}\ \bibnamefont
  {Cazalilla}},\ }\href {\doibase 10.1103/PhysRevA.80.063619} {\bibfield
  {journal} {\bibinfo  {journal} {Phys. Rev. A}\ }\textbf {\bibinfo {volume}
  {80}},\ \bibinfo {pages} {063619} (\bibinfo {year} {2009})}\BibitemShut
  {NoStop}%
\bibitem [{\citenamefont {Jaynes}(1957)}]{Jaynes57}%
  \BibitemOpen
  \bibfield  {author} {\bibinfo {author} {\bibfnamefont {E.~T.}\ \bibnamefont
  {Jaynes}},\ }\href {\doibase 10.1103/PhysRev.106.620} {\bibfield  {journal}
  {\bibinfo  {journal} {Phys. Rev.}\ }\textbf {\bibinfo {volume} {106}},\
  \bibinfo {pages} {620} (\bibinfo {year} {1957})}\BibitemShut {NoStop}%
\bibitem [{\citenamefont {Barthel}\ and\ \citenamefont
  {Schollw\"ock}(2008)}]{Barthel08}%
  \BibitemOpen
  \bibfield  {author} {\bibinfo {author} {\bibfnamefont {T.}~\bibnamefont
  {Barthel}}\ and\ \bibinfo {author} {\bibfnamefont {U.}~\bibnamefont
  {Schollw\"ock}},\ }\href {\doibase 10.1103/PhysRevLett.100.100601} {\bibfield
   {journal} {\bibinfo  {journal} {Phys. Rev. Lett.}\ }\textbf {\bibinfo
  {volume} {100}},\ \bibinfo {pages} {100601} (\bibinfo {year}
  {2008})}\BibitemShut {NoStop}%
\bibitem [{\citenamefont {Goldstein}\ and\ \citenamefont
  {Andrei}()}]{Goldstein14}%
  \BibitemOpen
  \bibfield  {author} {\bibinfo {author} {\bibfnamefont {G.}~\bibnamefont
  {Goldstein}}\ and\ \bibinfo {author} {\bibfnamefont {N.}~\bibnamefont
  {Andrei}},\ }\href@noop {} {\bibinfo  {journal} {arXiv:1405.4224}\
  }\BibitemShut {NoStop}%
\bibitem [{\citenamefont {Pozsgay}\ \emph {et~al.}(2014)\citenamefont
  {Pozsgay}, \citenamefont {Mesty\'an}, \citenamefont {Werner}, \citenamefont
  {Kormos}, \citenamefont {Zar\'and},\ and\ \citenamefont
  {Tak\'acs}}]{Pozsgay14}%
  \BibitemOpen
\bibfield  {journal} {  }\bibfield  {author} {\bibinfo {author} {\bibfnamefont
  {B.}~\bibnamefont {Pozsgay}}, \bibinfo {author} {\bibfnamefont
  {M.}~\bibnamefont {Mesty\'an}}, \bibinfo {author} {\bibfnamefont {M.~A.}\
  \bibnamefont {Werner}}, \bibinfo {author} {\bibfnamefont {M.}~\bibnamefont
  {Kormos}}, \bibinfo {author} {\bibfnamefont {G.}~\bibnamefont {Zar\'and}}, \
  and\ \bibinfo {author} {\bibfnamefont {G.}~\bibnamefont {Tak\'acs}},\ }\href
  {\doibase 10.1103/PhysRevLett.113.117203} {\bibfield  {journal} {\bibinfo
  {journal} {Phys. Rev. Lett.}\ }\textbf {\bibinfo {volume} {113}},\ \bibinfo
  {pages} {117203} (\bibinfo {year} {2014})}\BibitemShut {NoStop}%
\bibitem [{\citenamefont {Mierzejewski}\ \emph {et~al.}(2014)\citenamefont
  {Mierzejewski}, \citenamefont {Prelov\ifmmode~\check{s}\else \v{s}\fi{}ek},\
  and\ \citenamefont {Prosen}}]{Mierzejewski14}%
  \BibitemOpen
  \bibfield  {author} {\bibinfo {author} {\bibfnamefont {M.}~\bibnamefont
  {Mierzejewski}}, \bibinfo {author} {\bibfnamefont {P.}~\bibnamefont
  {Prelov\ifmmode~\check{s}\else \v{s}\fi{}ek}}, \ and\ \bibinfo {author}
  {\bibfnamefont {T.}~\bibnamefont {Prosen}},\ }\href {\doibase
  10.1103/PhysRevLett.113.020602} {\bibfield  {journal} {\bibinfo  {journal}
  {Phys. Rev. Lett.}\ }\textbf {\bibinfo {volume} {113}},\ \bibinfo {pages}
  {020602} (\bibinfo {year} {2014})}\BibitemShut {NoStop}%
\bibitem [{\citenamefont {Wouters}\ \emph {et~al.}(2014)\citenamefont
  {Wouters}, \citenamefont {De~Nardis}, \citenamefont {Brockmann},
  \citenamefont {Fioretto}, \citenamefont {Rigol},\ and\ \citenamefont
  {Caux}}]{Wouters14}%
  \BibitemOpen
  \bibfield  {author} {\bibinfo {author} {\bibfnamefont {B.}~\bibnamefont
  {Wouters}}, \bibinfo {author} {\bibfnamefont {J.}~\bibnamefont {De~Nardis}},
  \bibinfo {author} {\bibfnamefont {M.}~\bibnamefont {Brockmann}}, \bibinfo
  {author} {\bibfnamefont {D.}~\bibnamefont {Fioretto}}, \bibinfo {author}
  {\bibfnamefont {M.}~\bibnamefont {Rigol}}, \ and\ \bibinfo {author}
  {\bibfnamefont {J.-S.}\ \bibnamefont {Caux}},\ }\href {\doibase
  10.1103/PhysRevLett.113.117202} {\bibfield  {journal} {\bibinfo  {journal}
  {Phys. Rev. Lett.}\ }\textbf {\bibinfo {volume} {113}},\ \bibinfo {pages}
  {117202} (\bibinfo {year} {2014})}\BibitemShut {NoStop}%
\bibitem [{\citenamefont {Essler}\ \emph {et~al.}(2015)\citenamefont {Essler},
  \citenamefont {Mussardo},\ and\ \citenamefont {Panfil}}]{Muss14}%
  \BibitemOpen
  \bibfield  {author} {\bibinfo {author} {\bibfnamefont {F.~H.~L.}\
  \bibnamefont {Essler}}, \bibinfo {author} {\bibfnamefont {G.}~\bibnamefont
  {Mussardo}}, \ and\ \bibinfo {author} {\bibfnamefont {M.}~\bibnamefont
  {Panfil}},\ }\href {\doibase 10.1103/PhysRevA.91.051602} {\bibfield
  {journal} {\bibinfo  {journal} {Phys. Rev. A}\ }\textbf {\bibinfo {volume}
  {91}},\ \bibinfo {pages} {051602} (\bibinfo {year} {2015})}\BibitemShut
  {NoStop}%
\bibitem [{\citenamefont {Berges}\ \emph {et~al.}(2008)\citenamefont {Berges},
  \citenamefont {Rothkopf},\ and\ \citenamefont {Schmidt}}]{Berges08}%
  \BibitemOpen
  \bibfield  {author} {\bibinfo {author} {\bibfnamefont {J.}~\bibnamefont
  {Berges}}, \bibinfo {author} {\bibfnamefont {A.}~\bibnamefont {Rothkopf}}, \
  and\ \bibinfo {author} {\bibfnamefont {J.}~\bibnamefont {Schmidt}},\ }\href
  {\doibase 10.1103/PhysRevLett.101.041603} {\bibfield  {journal} {\bibinfo
  {journal} {Phys. Rev. Lett.}\ }\textbf {\bibinfo {volume} {101}},\ \bibinfo
  {pages} {041603} (\bibinfo {year} {2008})}\BibitemShut {NoStop}%
\bibitem [{\citenamefont {Nowak}\ \emph {et~al.}(2011)\citenamefont {Nowak},
  \citenamefont {Sexty},\ and\ \citenamefont {Gasenzer}}]{Nowak11}%
  \BibitemOpen
  \bibfield  {author} {\bibinfo {author} {\bibfnamefont {B.}~\bibnamefont
  {Nowak}}, \bibinfo {author} {\bibfnamefont {D.}~\bibnamefont {Sexty}}, \ and\
  \bibinfo {author} {\bibfnamefont {T.}~\bibnamefont {Gasenzer}},\ }\href
  {\doibase 10.1103/PhysRevB.84.020506} {\bibfield  {journal} {\bibinfo
  {journal} {Phys. Rev. B}\ }\textbf {\bibinfo {volume} {84}},\ \bibinfo
  {pages} {020506} (\bibinfo {year} {2011})}\BibitemShut {NoStop}%
\bibitem [{\citenamefont {Nowak}\ \emph {et~al.}(2014)\citenamefont {Nowak},
  \citenamefont {Schole},\ and\ \citenamefont {Gasenzer}}]{Nowak14}%
  \BibitemOpen
  \bibfield  {author} {\bibinfo {author} {\bibfnamefont {B.}~\bibnamefont
  {Nowak}}, \bibinfo {author} {\bibfnamefont {J.}~\bibnamefont {Schole}}, \
  and\ \bibinfo {author} {\bibfnamefont {T.}~\bibnamefont {Gasenzer}},\
  }\href@noop {} {\bibfield  {journal} {\bibinfo  {journal} {New J. Phys.}\
  }\textbf {\bibinfo {volume} {16}} (\bibinfo {year} {2014})}\BibitemShut
  {NoStop}%
\bibitem [{\citenamefont {Kolodrubetz}\ \emph {et~al.}(2012)\citenamefont
  {Kolodrubetz}, \citenamefont {Clark},\ and\ \citenamefont {Huse}}]{Huse12}%
  \BibitemOpen
  \bibfield  {author} {\bibinfo {author} {\bibfnamefont {M.}~\bibnamefont
  {Kolodrubetz}}, \bibinfo {author} {\bibfnamefont {B.~K.}\ \bibnamefont
  {Clark}}, \ and\ \bibinfo {author} {\bibfnamefont {D.~A.}\ \bibnamefont
  {Huse}},\ }\href {\doibase 10.1103/PhysRevLett.109.015701} {\bibfield
  {journal} {\bibinfo  {journal} {Phys. Rev. Lett.}\ }\textbf {\bibinfo
  {volume} {109}},\ \bibinfo {pages} {015701} (\bibinfo {year}
  {2012})}\BibitemShut {NoStop}%
\bibitem [{\citenamefont {Gambassi}\ and\ \citenamefont
  {Silva}()}]{Gambassi11}%
  \BibitemOpen
  \bibfield  {author} {\bibinfo {author} {\bibfnamefont {A.}~\bibnamefont
  {Gambassi}}\ and\ \bibinfo {author} {\bibfnamefont {A.}~\bibnamefont
  {Silva}},\ }\href@noop {} {\bibinfo  {journal} {arXiv:1106.2671}\
  }\BibitemShut {NoStop}%
\bibitem [{\citenamefont {Gambassi}\ and\ \citenamefont
  {Silva}(2012)}]{GambassiSilva12}%
  \BibitemOpen
\bibfield  {journal} {  }\bibfield  {author} {\bibinfo {author} {\bibfnamefont
  {A.}~\bibnamefont {Gambassi}}\ and\ \bibinfo {author} {\bibfnamefont
  {A.}~\bibnamefont {Silva}},\ }\href {\doibase 10.1103/PhysRevLett.109.250602}
  {\bibfield  {journal} {\bibinfo  {journal} {Phys. Rev. Lett.}\ }\textbf
  {\bibinfo {volume} {109}},\ \bibinfo {pages} {250602} (\bibinfo {year}
  {2012})}\BibitemShut {NoStop}%
\bibitem [{\citenamefont {Sotiriadis}\ \emph {et~al.}(2013)\citenamefont
  {Sotiriadis}, \citenamefont {Gambassi},\ and\ \citenamefont
  {Silva}}]{Sotiriadis13}%
  \BibitemOpen
  \bibfield  {author} {\bibinfo {author} {\bibfnamefont {S.}~\bibnamefont
  {Sotiriadis}}, \bibinfo {author} {\bibfnamefont {A.}~\bibnamefont
  {Gambassi}}, \ and\ \bibinfo {author} {\bibfnamefont {A.}~\bibnamefont
  {Silva}},\ }\href {\doibase 10.1103/PhysRevE.87.052129} {\bibfield  {journal}
  {\bibinfo  {journal} {Phys. Rev. E}\ }\textbf {\bibinfo {volume} {87}},\
  \bibinfo {pages} {052129} (\bibinfo {year} {2013})}\BibitemShut {NoStop}%
\bibitem [{\citenamefont {Dalla~Torre}\ \emph {et~al.}(2013)\citenamefont
  {Dalla~Torre}, \citenamefont {Demler},\ and\ \citenamefont
  {Polkovnikov}}]{Torre12b}%
  \BibitemOpen
  \bibfield  {author} {\bibinfo {author} {\bibfnamefont {E.~G.}\ \bibnamefont
  {Dalla~Torre}}, \bibinfo {author} {\bibfnamefont {E.}~\bibnamefont {Demler}},
  \ and\ \bibinfo {author} {\bibfnamefont {A.}~\bibnamefont {Polkovnikov}},\
  }\href {\doibase 10.1103/PhysRevLett.110.090404} {\bibfield  {journal}
  {\bibinfo  {journal} {Phys. Rev. Lett.}\ }\textbf {\bibinfo {volume} {110}},\
  \bibinfo {pages} {090404} (\bibinfo {year} {2013})}\BibitemShut {NoStop}%
\bibitem [{\citenamefont {Sciolla}\ and\ \citenamefont
  {Biroli}(2010)}]{Sciolla2010}%
  \BibitemOpen
  \bibfield  {author} {\bibinfo {author} {\bibfnamefont {B.}~\bibnamefont
  {Sciolla}}\ and\ \bibinfo {author} {\bibfnamefont {G.}~\bibnamefont
  {Biroli}},\ }\href {\doibase 10.1103/PhysRevLett.105.220401} {\bibfield
  {journal} {\bibinfo  {journal} {Phys. Rev. Lett.}\ }\textbf {\bibinfo
  {volume} {105}},\ \bibinfo {pages} {220401} (\bibinfo {year}
  {2010})}\BibitemShut {NoStop}%
\bibitem [{\citenamefont {Gambassi}\ and\ \citenamefont
  {Calabrese}(2011)}]{Gambassi2011}%
  \BibitemOpen
  \bibfield  {author} {\bibinfo {author} {\bibfnamefont {A.}~\bibnamefont
  {Gambassi}}\ and\ \bibinfo {author} {\bibfnamefont {P.}~\bibnamefont
  {Calabrese}},\ }\href@noop {} {\bibfield  {journal} {\bibinfo  {journal} {EPL
  (Europhysics Letters)}\ }\textbf {\bibinfo {volume} {95}},\ \bibinfo {pages}
  {66007} (\bibinfo {year} {2011})}\BibitemShut {NoStop}%
\bibitem [{\citenamefont {Sciolla}\ and\ \citenamefont
  {Biroli}(2011)}]{Sciolla11}%
  \BibitemOpen
  \bibfield  {author} {\bibinfo {author} {\bibfnamefont {B.}~\bibnamefont
  {Sciolla}}\ and\ \bibinfo {author} {\bibfnamefont {G.}~\bibnamefont
  {Biroli}},\ }\href@noop {} {\bibfield  {journal} {\bibinfo  {journal} {J.
  Stat. Mech.: Theor. Exp.}\ }\textbf {\bibinfo {volume} {2011}},\ \bibinfo
  {pages} {P11003} (\bibinfo {year} {2011})}\BibitemShut {NoStop}%
\bibitem [{\citenamefont {Sciolla}\ and\ \citenamefont
  {Biroli}(2013)}]{Sciolla2013}%
  \BibitemOpen
  \bibfield  {author} {\bibinfo {author} {\bibfnamefont {B.}~\bibnamefont
  {Sciolla}}\ and\ \bibinfo {author} {\bibfnamefont {G.}~\bibnamefont
  {Biroli}},\ }\href {\doibase 10.1103/PhysRevB.88.201110} {\bibfield
  {journal} {\bibinfo  {journal} {Phys. Rev. B}\ }\textbf {\bibinfo {volume}
  {88}},\ \bibinfo {pages} {201110(R)} (\bibinfo {year} {2013})}\BibitemShut
  {NoStop}%
\bibitem [{\citenamefont {Chandran}\ \emph {et~al.}(2013)\citenamefont
  {Chandran}, \citenamefont {Nanduri}, \citenamefont {Gubser},\ and\
  \citenamefont {Sondhi}}]{Sondhi2013}%
  \BibitemOpen
  \bibfield  {author} {\bibinfo {author} {\bibfnamefont {A.}~\bibnamefont
  {Chandran}}, \bibinfo {author} {\bibfnamefont {A.}~\bibnamefont {Nanduri}},
  \bibinfo {author} {\bibfnamefont {S.~S.}\ \bibnamefont {Gubser}}, \ and\
  \bibinfo {author} {\bibfnamefont {S.~L.}\ \bibnamefont {Sondhi}},\ }\href
  {\doibase 10.1103/PhysRevB.88.024306} {\bibfield  {journal} {\bibinfo
  {journal} {Phys. Rev. B}\ }\textbf {\bibinfo {volume} {88}},\ \bibinfo
  {pages} {024306} (\bibinfo {year} {2013})}\BibitemShut {NoStop}%
\bibitem [{\citenamefont {{Smacchia}}\ \emph {et~al.}()\citenamefont
  {{Smacchia}}, \citenamefont {{Knap}}, \citenamefont {{Demler}},\ and\
  \citenamefont {{Silva}}}]{Smacchia}%
  \BibitemOpen
  \bibfield  {author} {\bibinfo {author} {\bibfnamefont {P.}~\bibnamefont
  {{Smacchia}}}, \bibinfo {author} {\bibfnamefont {M.}~\bibnamefont {{Knap}}},
  \bibinfo {author} {\bibfnamefont {E.}~\bibnamefont {{Demler}}}, \ and\
  \bibinfo {author} {\bibfnamefont {A.}~\bibnamefont {{Silva}}},\ }\href@noop
  {} {\bibinfo  {journal} {arXiv:1409.1883}\ }\BibitemShut {NoStop}%
\bibitem [{\citenamefont {Eckstein}\ \emph {et~al.}(2009)\citenamefont
  {Eckstein}, \citenamefont {Kollar},\ and\ \citenamefont {Werner}}]{Werner09}%
  \BibitemOpen
\bibfield  {journal} {  }\bibfield  {author} {\bibinfo {author} {\bibfnamefont
  {M.}~\bibnamefont {Eckstein}}, \bibinfo {author} {\bibfnamefont
  {M.}~\bibnamefont {Kollar}}, \ and\ \bibinfo {author} {\bibfnamefont
  {P.}~\bibnamefont {Werner}},\ }\href {\doibase
  10.1103/PhysRevLett.103.056403} {\bibfield  {journal} {\bibinfo  {journal}
  {Phys. Rev. Lett.}\ }\textbf {\bibinfo {volume} {103}},\ \bibinfo {pages}
  {056403} (\bibinfo {year} {2009})}\BibitemShut {NoStop}%
\bibitem [{\citenamefont {Schir\'o}\ and\ \citenamefont
  {Fabrizio}(2010)}]{Schiro2010}%
  \BibitemOpen
  \bibfield  {author} {\bibinfo {author} {\bibfnamefont {M.}~\bibnamefont
  {Schir\'o}}\ and\ \bibinfo {author} {\bibfnamefont {M.}~\bibnamefont
  {Fabrizio}},\ }\href {\doibase 10.1103/PhysRevLett.105.076401} {\bibfield
  {journal} {\bibinfo  {journal} {Phys. Rev. Lett.}\ }\textbf {\bibinfo
  {volume} {105}},\ \bibinfo {pages} {076401} (\bibinfo {year}
  {2010})}\BibitemShut {NoStop}%
\bibitem [{\citenamefont {{Franchini}}\ \emph {et~al.}()\citenamefont
  {{Franchini}}, \citenamefont {{Gromov}}, \citenamefont {{Kulkarni}},\ and\
  \citenamefont {{Trombettoni}}}]{Franchini2014}%
  \BibitemOpen
  \bibfield  {author} {\bibinfo {author} {\bibfnamefont {F.}~\bibnamefont
  {{Franchini}}}, \bibinfo {author} {\bibfnamefont {A.}~\bibnamefont
  {{Gromov}}}, \bibinfo {author} {\bibfnamefont {M.}~\bibnamefont
  {{Kulkarni}}}, \ and\ \bibinfo {author} {\bibfnamefont {A.}~\bibnamefont
  {{Trombettoni}}},\ }\href@noop {} {\bibinfo  {journal} {arXiv:1408.3618}\
  }\BibitemShut {NoStop}%
\bibitem [{\citenamefont {Janssen}\ \emph {et~al.}(1989)\citenamefont
  {Janssen}, \citenamefont {Schaub},\ and\ \citenamefont
  {Schmittmann}}]{Janssen89}%
  \BibitemOpen
\bibfield  {journal} {  }\bibfield  {author} {\bibinfo {author} {\bibfnamefont
  {H.~K.}\ \bibnamefont {Janssen}}, \bibinfo {author} {\bibfnamefont
  {B.}~\bibnamefont {Schaub}}, \ and\ \bibinfo {author} {\bibfnamefont
  {B.}~\bibnamefont {Schmittmann}},\ }\href@noop {} {\bibfield  {journal}
  {\bibinfo  {journal} {Z. Phys. B}\ }\textbf {\bibinfo {volume} {73}},\
  \bibinfo {pages} {539} (\bibinfo {year} {1989})}\BibitemShut {NoStop}%
\bibitem [{\citenamefont {Calabrese}\ and\ \citenamefont
  {Gambassi}(2005)}]{Gambassi05}%
  \BibitemOpen
  \bibfield  {author} {\bibinfo {author} {\bibfnamefont {P.}~\bibnamefont
  {Calabrese}}\ and\ \bibinfo {author} {\bibfnamefont {A.}~\bibnamefont
  {Gambassi}},\ }\href@noop {} {\bibfield  {journal} {\bibinfo  {journal} {J.
  Phys. A: Math. Gen.}\ }\textbf {\bibinfo {volume} {38}},\ \bibinfo {pages}
  {R133} (\bibinfo {year} {2005})}\BibitemShut {NoStop}%
\bibitem [{\citenamefont {Bonart}\ \emph {et~al.}(2012)\citenamefont {Bonart},
  \citenamefont {Cugliandolo},\ and\ \citenamefont {Gambassi}}]{Bonart12}%
  \BibitemOpen
  \bibfield  {author} {\bibinfo {author} {\bibfnamefont {J.}~\bibnamefont
  {Bonart}}, \bibinfo {author} {\bibfnamefont {L.~F.}\ \bibnamefont
  {Cugliandolo}}, \ and\ \bibinfo {author} {\bibfnamefont {A.}~\bibnamefont
  {Gambassi}},\ }\href@noop {} {\bibfield  {journal} {\bibinfo  {journal} {J.
  Stat. Mech.: Theor. Exp.}\ }\textbf {\bibinfo {volume} {2012}},\ \bibinfo
  {pages} {P01014} (\bibinfo {year} {2012})}\BibitemShut {NoStop}%
\bibitem [{\citenamefont {Marcuzzi}\ \emph {et~al.}(2012)\citenamefont
  {Marcuzzi}, \citenamefont {Gambassi},\ and\ \citenamefont
  {Pleimling}}]{Marcuzzi12}%
  \BibitemOpen
  \bibfield  {author} {\bibinfo {author} {\bibfnamefont {M.}~\bibnamefont
  {Marcuzzi}}, \bibinfo {author} {\bibfnamefont {A.}~\bibnamefont {Gambassi}},
  \ and\ \bibinfo {author} {\bibfnamefont {M.}~\bibnamefont {Pleimling}},\
  }\href@noop {} {\bibfield  {journal} {\bibinfo  {journal} {EPL (Europhysics
  Letters)}\ }\textbf {\bibinfo {volume} {100}},\ \bibinfo {pages} {46004}
  (\bibinfo {year} {2012})}\BibitemShut {NoStop}%
\bibitem [{\citenamefont {Hackl}\ \emph {et~al.}(2009)\citenamefont {Hackl},
  \citenamefont {Roosen}, \citenamefont {Kehrein},\ and\ \citenamefont
  {Hofstetter}}]{Hackl2009}%
  \BibitemOpen
  \bibfield  {author} {\bibinfo {author} {\bibfnamefont {A.}~\bibnamefont
  {Hackl}}, \bibinfo {author} {\bibfnamefont {D.}~\bibnamefont {Roosen}},
  \bibinfo {author} {\bibfnamefont {S.}~\bibnamefont {Kehrein}}, \ and\
  \bibinfo {author} {\bibfnamefont {W.}~\bibnamefont {Hofstetter}},\ }\href
  {\doibase 10.1103/PhysRevLett.102.196601} {\bibfield  {journal} {\bibinfo
  {journal} {Phys. Rev. Lett.}\ }\textbf {\bibinfo {volume} {102}},\ \bibinfo
  {pages} {196601} (\bibinfo {year} {2009})}\BibitemShut {NoStop}%
\bibitem [{\citenamefont {Pletyukhov}\ \emph {et~al.}(2010)\citenamefont
  {Pletyukhov}, \citenamefont {Schuricht},\ and\ \citenamefont
  {Schoeller}}]{Pletyukhov2010}%
  \BibitemOpen
  \bibfield  {author} {\bibinfo {author} {\bibfnamefont {M.}~\bibnamefont
  {Pletyukhov}}, \bibinfo {author} {\bibfnamefont {D.}~\bibnamefont
  {Schuricht}}, \ and\ \bibinfo {author} {\bibfnamefont {H.}~\bibnamefont
  {Schoeller}},\ }\href {\doibase 10.1103/PhysRevLett.104.106801} {\bibfield
  {journal} {\bibinfo  {journal} {Phys. Rev. Lett.}\ }\textbf {\bibinfo
  {volume} {104}},\ \bibinfo {pages} {106801} (\bibinfo {year}
  {2010})}\BibitemShut {NoStop}%
\bibitem [{\citenamefont {Gagel}\ \emph {et~al.}(2014)\citenamefont {Gagel},
  \citenamefont {Orth},\ and\ \citenamefont {Schmalian}}]{Gagel14}%
  \BibitemOpen
  \bibfield  {author} {\bibinfo {author} {\bibfnamefont {P.}~\bibnamefont
  {Gagel}}, \bibinfo {author} {\bibfnamefont {P.~P.}\ \bibnamefont {Orth}}, \
  and\ \bibinfo {author} {\bibfnamefont {J.}~\bibnamefont {Schmalian}},\ }\href
  {\doibase 10.1103/PhysRevLett.113.220401} {\bibfield  {journal} {\bibinfo
  {journal} {Phys. Rev. Lett.}\ }\textbf {\bibinfo {volume} {113}},\ \bibinfo
  {pages} {220401} (\bibinfo {year} {2014})}\BibitemShut {NoStop}%
\bibitem [{\citenamefont {Buchhold}\ and\ \citenamefont
  {Diehl}()}]{Buchhold14}%
  \BibitemOpen
  \bibfield  {author} {\bibinfo {author} {\bibfnamefont {M.}~\bibnamefont
  {Buchhold}}\ and\ \bibinfo {author} {\bibfnamefont {S.}~\bibnamefont
  {Diehl}},\ }\href@noop {} {\bibinfo  {journal} {arXiv:1404.3740}\
  }\BibitemShut {NoStop}%
\bibitem [{\citenamefont {Diehl}(1986)}]{Diehl86}%
  \BibitemOpen
\bibfield  {journal} {  }\bibfield  {author} {\bibinfo {author} {\bibfnamefont
  {H.~W.}\ \bibnamefont {Diehl}},\ }in\ \href@noop {} {\emph {\bibinfo
  {booktitle} {Phase Transitions and Critical Phenomena}}},\ Vol.~\bibinfo
  {volume} {10},\ \bibinfo {editor} {edited by\ \bibinfo {editor}
  {\bibfnamefont {C.}~\bibnamefont {Domb}}\ and\ \bibinfo {editor}
  {\bibfnamefont {J.~L.}\ \bibnamefont {Lebowitz}}}\ (\bibinfo  {publisher}
  {Academic Press, London},\ \bibinfo {year} {1986})\BibitemShut {NoStop}%
\bibitem [{\citenamefont {Diehl}(1997)}]{Diehl97}%
  \BibitemOpen
  \bibfield  {author} {\bibinfo {author} {\bibfnamefont {H.~W.}\ \bibnamefont
  {Diehl}},\ }\href@noop {} {\bibfield  {journal} {\bibinfo  {journal} {Int. J.
  Mod. Phys. B}\ }\textbf {\bibinfo {volume} {11}},\ \bibinfo {pages} {3503}
  (\bibinfo {year} {1997})}\BibitemShut {NoStop}%
\bibitem [{\citenamefont {Pleimling}(2004)}]{Pleimling04}%
  \BibitemOpen
  \bibfield  {author} {\bibinfo {author} {\bibfnamefont {M.}~\bibnamefont
  {Pleimling}},\ }\href@noop {} {\bibfield  {journal} {\bibinfo  {journal} {J.
  Phys. A: Math. Gen.}\ }\textbf {\bibinfo {volume} {37}},\ \bibinfo {pages}
  {R79} (\bibinfo {year} {2004})}\BibitemShut {NoStop}%
\bibitem [{\citenamefont {Mahan}(2000)}]{Mahan}%
  \BibitemOpen
  \bibfield  {author} {\bibinfo {author} {\bibfnamefont {G.}~\bibnamefont
  {Mahan}},\ }\href@noop {} {\emph {\bibinfo {title} {Many-Particle
  Physics}}},\ Physics of Solids and Liquids\ (\bibinfo  {publisher}
  {Springer},\ \bibinfo {year} {2000})\BibitemShut {NoStop}%
\bibitem [{\citenamefont {Kamenev}(2011)}]{Kamenevbook}%
  \BibitemOpen
  \bibfield  {author} {\bibinfo {author} {\bibfnamefont {A.}~\bibnamefont
  {Kamenev}},\ }\href@noop {} {\emph {\bibinfo {title} {Field Theory of
  Non-Equilibrium Systems}}}\ (\bibinfo  {publisher} {Cambridge University
  Press},\ \bibinfo {year} {2011})\BibitemShut {NoStop}%
\bibitem [{\citenamefont {Calabrese}\ and\ \citenamefont
  {Cardy}(2007)}]{Calabrese07}%
  \BibitemOpen
  \bibfield  {author} {\bibinfo {author} {\bibfnamefont {P.}~\bibnamefont
  {Calabrese}}\ and\ \bibinfo {author} {\bibfnamefont {J.}~\bibnamefont
  {Cardy}},\ }\href {\doibase 10.1088/1742-5468/2007/06/P06008} {\bibfield
  {journal} {\bibinfo  {journal} {J. Stat. Mech.: Theor. Exp.}\ }\textbf
  {\bibinfo {volume} {2007}},\ \bibinfo {pages} {P06008} (\bibinfo {year}
  {2007})}\BibitemShut {NoStop}%
\bibitem [{\citenamefont {Marcuzzi}\ and\ \citenamefont
  {Gambassi}(2014)}]{PhysRevB.89.134307}%
  \BibitemOpen
  \bibfield  {author} {\bibinfo {author} {\bibfnamefont {M.}~\bibnamefont
  {Marcuzzi}}\ and\ \bibinfo {author} {\bibfnamefont {A.}~\bibnamefont
  {Gambassi}},\ }\href {\doibase 10.1103/PhysRevB.89.134307} {\bibfield
  {journal} {\bibinfo  {journal} {Phys. Rev. B}\ }\textbf {\bibinfo {volume}
  {89}},\ \bibinfo {pages} {134307} (\bibinfo {year} {2014})}\BibitemShut
  {NoStop}%
\bibitem [{\citenamefont {Sachdev}(2011)}]{Sachdev11}%
  \BibitemOpen
  \bibfield  {author} {\bibinfo {author} {\bibfnamefont {S.}~\bibnamefont
  {Sachdev}},\ }\href@noop {} {\emph {\bibinfo {title} {Quantum Phase
  Transitions}}},\ \bibinfo {edition} {2nd}\ ed.\ (\bibinfo  {publisher}
  {Cambridge University Press},\ \bibinfo {year} {2011})\BibitemShut {NoStop}%
\bibitem [{\citenamefont {Sondhi}\ \emph {et~al.}(1997)\citenamefont {Sondhi},
  \citenamefont {Girvin}, \citenamefont {Carini},\ and\ \citenamefont
  {Shahar}}]{Sondhi97}%
  \BibitemOpen
  \bibfield  {author} {\bibinfo {author} {\bibfnamefont {S.~L.}\ \bibnamefont
  {Sondhi}}, \bibinfo {author} {\bibfnamefont {S.~M.}\ \bibnamefont {Girvin}},
  \bibinfo {author} {\bibfnamefont {J.~P.}\ \bibnamefont {Carini}}, \ and\
  \bibinfo {author} {\bibfnamefont {D.}~\bibnamefont {Shahar}},\ }\href
  {\doibase 10.1103/RevModPhys.69.315} {\bibfield  {journal} {\bibinfo
  {journal} {Rev. Mod. Phys.}\ }\textbf {\bibinfo {volume} {69}},\ \bibinfo
  {pages} {315} (\bibinfo {year} {1997})}\BibitemShut {NoStop}%
\bibitem [{\citenamefont {Fisher}\ and\ \citenamefont
  {Hohenberg}(1988{\natexlab{a}})}]{Fisher88}%
  \BibitemOpen
  \bibfield  {author} {\bibinfo {author} {\bibfnamefont {D.~S.}\ \bibnamefont
  {Fisher}}\ and\ \bibinfo {author} {\bibfnamefont {P.~C.}\ \bibnamefont
  {Hohenberg}},\ }\href {\doibase 10.1103/PhysRevB.37.4936} {\bibfield
  {journal} {\bibinfo  {journal} {Phys. Rev. B}\ }\textbf {\bibinfo {volume}
  {37}},\ \bibinfo {pages} {4936} (\bibinfo {year}
  {1988}{\natexlab{a}})}\BibitemShut {NoStop}%
\bibitem [{\citenamefont {Wilson}\ and\ \citenamefont
  {Kogut}(1974)}]{Wilson74}%
  \BibitemOpen
  \bibfield  {author} {\bibinfo {author} {\bibfnamefont {K.}~\bibnamefont
  {Wilson}}\ and\ \bibinfo {author} {\bibfnamefont {J.}~\bibnamefont {Kogut}},\
  }\href {\doibase 10.1016/0370-1573(74)90023-4} {\bibfield  {journal}
  {\bibinfo  {journal} {Physics Reports}\ }\textbf {\bibinfo {volume} {12}},\
  \bibinfo {pages} {75} (\bibinfo {year} {1974})}\BibitemShut {NoStop}%
\bibitem [{\citenamefont {Fisher}(1998)}]{Fisher98}%
  \BibitemOpen
  \bibfield  {author} {\bibinfo {author} {\bibfnamefont {M.~E.}\ \bibnamefont
  {Fisher}},\ }\href {\doibase 10.1103/RevModPhys.70.653} {\bibfield  {journal}
  {\bibinfo  {journal} {Rev. Mod. Phys.}\ }\textbf {\bibinfo {volume} {70}},\
  \bibinfo {pages} {653} (\bibinfo {year} {1998})}\BibitemShut {NoStop}%
\bibitem [{\citenamefont {Mitra}(2012)}]{Mitra12b}%
  \BibitemOpen
  \bibfield  {author} {\bibinfo {author} {\bibfnamefont {A.}~\bibnamefont
  {Mitra}},\ }\href {\doibase 10.1103/PhysRevLett.109.260601} {\bibfield
  {journal} {\bibinfo  {journal} {Phys. Rev. Lett.}\ }\textbf {\bibinfo
  {volume} {109}},\ \bibinfo {pages} {260601} (\bibinfo {year}
  {2012})}\BibitemShut {NoStop}%
\bibitem [{\citenamefont {Mitra}\ and\ \citenamefont
  {Millis}(2008)}]{Mitra08a}%
  \BibitemOpen
  \bibfield  {author} {\bibinfo {author} {\bibfnamefont {A.}~\bibnamefont
  {Mitra}}\ and\ \bibinfo {author} {\bibfnamefont {A.~J.}\ \bibnamefont
  {Millis}},\ }\href {\doibase 10.1103/PhysRevB.77.220404} {\bibfield
  {journal} {\bibinfo  {journal} {Phys. Rev. B}\ }\textbf {\bibinfo {volume}
  {77}},\ \bibinfo {pages} {220404} (\bibinfo {year} {2008})}\BibitemShut
  {NoStop}%
\bibitem [{\citenamefont {Sarkar}\ \emph {et~al.}(2014)\citenamefont {Sarkar},
  \citenamefont {Sensarma},\ and\ \citenamefont {Sengupta}}]{Sengupta14}%
  \BibitemOpen
  \bibfield  {author} {\bibinfo {author} {\bibfnamefont {S.~D.}\ \bibnamefont
  {Sarkar}}, \bibinfo {author} {\bibfnamefont {R.}~\bibnamefont {Sensarma}}, \
  and\ \bibinfo {author} {\bibfnamefont {K.}~\bibnamefont {Sengupta}},\
  }\href@noop {} {\bibfield  {journal} {\bibinfo  {journal} {J. Phys.: Cond.
  Matt.}\ }\textbf {\bibinfo {volume} {26}},\ \bibinfo {pages} {325602}
  (\bibinfo {year} {2014})}\BibitemShut {NoStop}%
\bibitem [{\citenamefont {Cardy}(1988)}]{FiniteSize}%
  \BibitemOpen
  \bibinfo {editor} {\bibfnamefont {J.~L.}\ \bibnamefont {Cardy}},\ ed.,\
  \href@noop {} {\emph {\bibinfo {title} {Finite-Size Scaling}}},\ \bibinfo
  {series} {Current Physics Sources and Comments}, Vol.~\bibinfo {volume} {2}\
  (\bibinfo  {publisher} {Elsevier},\ \bibinfo {year} {1988})\BibitemShut
  {NoStop}%
\bibitem [{\citenamefont {Foini}\ \emph
  {et~al.}(2011{\natexlab{a}})\citenamefont {Foini}, \citenamefont
  {Cugliandolo},\ and\ \citenamefont {Gambassi}}]{PhysRevB.84.212404}%
  \BibitemOpen
  \bibfield  {author} {\bibinfo {author} {\bibfnamefont {L.}~\bibnamefont
  {Foini}}, \bibinfo {author} {\bibfnamefont {L.~F.}\ \bibnamefont
  {Cugliandolo}}, \ and\ \bibinfo {author} {\bibfnamefont {A.}~\bibnamefont
  {Gambassi}},\ }\href {\doibase 10.1103/PhysRevB.84.212404} {\bibfield
  {journal} {\bibinfo  {journal} {Phys. Rev. B}\ }\textbf {\bibinfo {volume}
  {84}},\ \bibinfo {pages} {212404} (\bibinfo {year}
  {2011}{\natexlab{a}})}\BibitemShut {NoStop}%
\bibitem [{\citenamefont {Foini}\ \emph
  {et~al.}(2011{\natexlab{b}})\citenamefont {Foini}, \citenamefont
  {Cugliandolo},\ and\ \citenamefont {Gambassi}}]{Foini11}%
  \BibitemOpen
  \bibfield  {author} {\bibinfo {author} {\bibfnamefont {L.}~\bibnamefont
  {Foini}}, \bibinfo {author} {\bibfnamefont {L.~F.}\ \bibnamefont
  {Cugliandolo}}, \ and\ \bibinfo {author} {\bibfnamefont {A.}~\bibnamefont
  {Gambassi}},\ }\href {\doibase 10.1088/1742-5468/2012/09/P09011} {\bibfield
  {journal} {\bibinfo  {journal} {J. Stat. Mech.: Theor. Exp.}\ }\textbf
  {\bibinfo {volume} {2011}},\ \bibinfo {pages} {P09011} (\bibinfo {year}
  {2011}{\natexlab{b}})}\BibitemShut {NoStop}%
\bibitem [{\citenamefont {Mitra}\ and\ \citenamefont
  {Giamarchi}(2011)}]{Mitra11}%
  \BibitemOpen
  \bibfield  {author} {\bibinfo {author} {\bibfnamefont {A.}~\bibnamefont
  {Mitra}}\ and\ \bibinfo {author} {\bibfnamefont {T.}~\bibnamefont
  {Giamarchi}},\ }\href {\doibase 10.1103/PhysRevLett.107.150602} {\bibfield
  {journal} {\bibinfo  {journal} {Phys. Rev. Lett.}\ }\textbf {\bibinfo
  {volume} {107}},\ \bibinfo {pages} {150602} (\bibinfo {year}
  {2011})}\BibitemShut {NoStop}%
\bibitem [{\citenamefont {Mitra}\ and\ \citenamefont
  {Giamarchi}(2012)}]{Mitra12a}%
  \BibitemOpen
  \bibfield  {author} {\bibinfo {author} {\bibfnamefont {A.}~\bibnamefont
  {Mitra}}\ and\ \bibinfo {author} {\bibfnamefont {T.}~\bibnamefont
  {Giamarchi}},\ }\href {\doibase 10.1103/PhysRevB.85.075117} {\bibfield
  {journal} {\bibinfo  {journal} {Phys. Rev. B}\ }\textbf {\bibinfo {volume}
  {85}},\ \bibinfo {pages} {075117} (\bibinfo {year} {2012})}\BibitemShut
  {NoStop}%
\bibitem [{\citenamefont {Tavora}\ and\ \citenamefont
  {Mitra}(2013)}]{Tavora13}%
  \BibitemOpen
  \bibfield  {author} {\bibinfo {author} {\bibfnamefont {M.}~\bibnamefont
  {Tavora}}\ and\ \bibinfo {author} {\bibfnamefont {A.}~\bibnamefont {Mitra}},\
  }\href {\doibase 10.1103/PhysRevB.88.115144} {\bibfield  {journal} {\bibinfo
  {journal} {Phys. Rev. B}\ }\textbf {\bibinfo {volume} {88}},\ \bibinfo
  {pages} {115144} (\bibinfo {year} {2013})}\BibitemShut {NoStop}%
\bibitem [{\citenamefont {Lux}\ \emph {et~al.}(2014)\citenamefont {Lux},
  \citenamefont {M\"uller}, \citenamefont {Mitra},\ and\ \citenamefont
  {Rosch}}]{Lux13}%
  \BibitemOpen
  \bibfield  {author} {\bibinfo {author} {\bibfnamefont {J.}~\bibnamefont
  {Lux}}, \bibinfo {author} {\bibfnamefont {J.}~\bibnamefont {M\"uller}},
  \bibinfo {author} {\bibfnamefont {A.}~\bibnamefont {Mitra}}, \ and\ \bibinfo
  {author} {\bibfnamefont {A.}~\bibnamefont {Rosch}},\ }\href {\doibase
  10.1103/PhysRevA.89.053608} {\bibfield  {journal} {\bibinfo  {journal} {Phys.
  Rev. A}\ }\textbf {\bibinfo {volume} {89}},\ \bibinfo {pages} {053608}
  (\bibinfo {year} {2014})}\BibitemShut {NoStop}%
\bibitem [{\citenamefont {Chiocchetta}\ \emph {et~al.}(2015)\citenamefont
  {Chiocchetta}, \citenamefont {Tavora}, \citenamefont {Gambassi},\ and\
  \citenamefont {Mitra}}]{Chiocchetta14}%
  \BibitemOpen
  \bibfield  {author} {\bibinfo {author} {\bibfnamefont {A.}~\bibnamefont
  {Chiocchetta}}, \bibinfo {author} {\bibfnamefont {M.}~\bibnamefont {Tavora}},
  \bibinfo {author} {\bibfnamefont {A.}~\bibnamefont {Gambassi}}, \ and\
  \bibinfo {author} {\bibfnamefont {A.}~\bibnamefont {Mitra}},\ }\href@noop {}
  {\bibfield  {journal} {\bibinfo  {journal} {in preparation}\ } (\bibinfo
  {year} {2015})}\BibitemShut {NoStop}%
\bibitem [{\citenamefont {Calabrese}\ and\ \citenamefont
  {Cardy}(2006)}]{Calabrese06}%
  \BibitemOpen
  \bibfield  {author} {\bibinfo {author} {\bibfnamefont {P.}~\bibnamefont
  {Calabrese}}\ and\ \bibinfo {author} {\bibfnamefont {J.}~\bibnamefont
  {Cardy}},\ }\href {\doibase 10.1103/PhysRevLett.96.136801} {\bibfield
  {journal} {\bibinfo  {journal} {Phys. Rev. Lett.}\ }\textbf {\bibinfo
  {volume} {96}},\ \bibinfo {pages} {136801} (\bibinfo {year}
  {2006})}\BibitemShut {NoStop}%
\bibitem [{\citenamefont {Karl}\ \emph {et~al.}(2013)\citenamefont {Karl},
  \citenamefont {Nowak},\ and\ \citenamefont {Gasenzer}}]{PhysRevA.88.063615}%
  \BibitemOpen
  \bibfield  {author} {\bibinfo {author} {\bibfnamefont {M.}~\bibnamefont
  {Karl}}, \bibinfo {author} {\bibfnamefont {B.}~\bibnamefont {Nowak}}, \ and\
  \bibinfo {author} {\bibfnamefont {T.}~\bibnamefont {Gasenzer}},\ }\href
  {\doibase 10.1103/PhysRevA.88.063615} {\bibfield  {journal} {\bibinfo
  {journal} {Phys. Rev. A}\ }\textbf {\bibinfo {volume} {88}},\ \bibinfo
  {pages} {063615} (\bibinfo {year} {2013})}\BibitemShut {NoStop}%
\bibitem [{\citenamefont {Bakr}\ \emph {et~al.}(2009)\citenamefont {Bakr},
  \citenamefont {Gillen}, \citenamefont {Peng}, \citenamefont {Folling},\ and\
  \citenamefont {Greiner}}]{Bakr2009}%
  \BibitemOpen
  \bibfield  {author} {\bibinfo {author} {\bibfnamefont {W.~S.}\ \bibnamefont
  {Bakr}}, \bibinfo {author} {\bibfnamefont {J.~I.}\ \bibnamefont {Gillen}},
  \bibinfo {author} {\bibfnamefont {A.}~\bibnamefont {Peng}}, \bibinfo {author}
  {\bibfnamefont {S.}~\bibnamefont {Folling}}, \ and\ \bibinfo {author}
  {\bibfnamefont {M.}~\bibnamefont {Greiner}},\ }\href@noop {} {\bibfield
  {journal} {\bibinfo  {journal} {Nature}\ }\textbf {\bibinfo {volume} {462}},\
  \bibinfo {pages} {74} (\bibinfo {year} {2009})}\BibitemShut {NoStop}%
\bibitem [{\citenamefont {Bakr}\ \emph {et~al.}(2010)\citenamefont {Bakr},
  \citenamefont {Peng}, \citenamefont {Tai}, \citenamefont {Ma}, \citenamefont
  {Simon}, \citenamefont {Gillen}, \citenamefont {Fölling}, \citenamefont
  {Pollet},\ and\ \citenamefont {Greiner}}]{Bakr2010}%
  \BibitemOpen
  \bibfield  {author} {\bibinfo {author} {\bibfnamefont {W.~S.}\ \bibnamefont
  {Bakr}}, \bibinfo {author} {\bibfnamefont {A.}~\bibnamefont {Peng}}, \bibinfo
  {author} {\bibfnamefont {M.~E.}\ \bibnamefont {Tai}}, \bibinfo {author}
  {\bibfnamefont {R.}~\bibnamefont {Ma}}, \bibinfo {author} {\bibfnamefont
  {J.}~\bibnamefont {Simon}}, \bibinfo {author} {\bibfnamefont {J.~I.}\
  \bibnamefont {Gillen}}, \bibinfo {author} {\bibfnamefont {S.}~\bibnamefont
  {Fölling}}, \bibinfo {author} {\bibfnamefont {L.}~\bibnamefont {Pollet}}, \
  and\ \bibinfo {author} {\bibfnamefont {M.}~\bibnamefont {Greiner}},\ }\href
  {\doibase 10.1126/science.1192368} {\bibfield  {journal} {\bibinfo  {journal}
  {Science}\ }\textbf {\bibinfo {volume} {329}},\ \bibinfo {pages} {547}
  (\bibinfo {year} {2010})}\BibitemShut {NoStop}%
\bibitem [{\citenamefont {Betz}\ \emph {et~al.}(2011)\citenamefont {Betz},
  \citenamefont {Manz}, \citenamefont {B\"ucker}, \citenamefont {Berrada},
  \citenamefont {Koller}, \citenamefont {Kazakov}, \citenamefont {Mazets},
  \citenamefont {Stimming}, \citenamefont {Perrin}, \citenamefont {Schumm},\
  and\ \citenamefont {Schmiedmayer}}]{Betz2011}%
  \BibitemOpen
  \bibfield  {author} {\bibinfo {author} {\bibfnamefont {T.}~\bibnamefont
  {Betz}}, \bibinfo {author} {\bibfnamefont {S.}~\bibnamefont {Manz}}, \bibinfo
  {author} {\bibfnamefont {R.}~\bibnamefont {B\"ucker}}, \bibinfo {author}
  {\bibfnamefont {T.}~\bibnamefont {Berrada}}, \bibinfo {author} {\bibfnamefont
  {C.}~\bibnamefont {Koller}}, \bibinfo {author} {\bibfnamefont
  {G.}~\bibnamefont {Kazakov}}, \bibinfo {author} {\bibfnamefont
  {I.}~\bibnamefont {Mazets}}, \bibinfo {author} {\bibfnamefont {H.-P.}\
  \bibnamefont {Stimming}}, \bibinfo {author} {\bibfnamefont {A.}~\bibnamefont
  {Perrin}}, \bibinfo {author} {\bibfnamefont {T.}~\bibnamefont {Schumm}}, \
  and\ \bibinfo {author} {\bibfnamefont {J.}~\bibnamefont {Schmiedmayer}},\
  }\href {\doibase 10.1103/PhysRevLett.106.020407} {\bibfield  {journal}
  {\bibinfo  {journal} {Phys. Rev. Lett.}\ }\textbf {\bibinfo {volume} {106}},\
  \bibinfo {pages} {020407} (\bibinfo {year} {2011})}\BibitemShut {NoStop}%
\bibitem [{\citenamefont {{Larr{\'e}}}\ and\ \citenamefont
  {{Carusotto}}()}]{Larre2014}%
  \BibitemOpen
  \bibfield  {author} {\bibinfo {author} {\bibfnamefont {P.-{\'E}.}\
  \bibnamefont {{Larr{\'e}}}}\ and\ \bibinfo {author} {\bibfnamefont
  {I.}~\bibnamefont {{Carusotto}}},\ }\href@noop {} {\bibinfo  {journal}
  {arXiv:1412.5405}\ }\BibitemShut {NoStop}%
\bibitem [{\citenamefont {Gorshkov}\ \emph {et~al.}(2010)\citenamefont
  {Gorshkov}, \citenamefont {Hermele}, \citenamefont {Gurarie}, \citenamefont
  {Xu}, \citenamefont {Julienne}, \citenamefont {Ye}, \citenamefont {Zoller},
  \citenamefont {Demler}, \citenamefont {Lukin},\ and\ \citenamefont
  {Rey}}]{Gorshkov2010}%
  \BibitemOpen
\bibfield  {journal} {  }\bibfield  {author} {\bibinfo {author} {\bibfnamefont
  {A.~V.}\ \bibnamefont {Gorshkov}}, \bibinfo {author} {\bibfnamefont
  {M.}~\bibnamefont {Hermele}}, \bibinfo {author} {\bibfnamefont
  {V.}~\bibnamefont {Gurarie}}, \bibinfo {author} {\bibfnamefont
  {C.}~\bibnamefont {Xu}}, \bibinfo {author} {\bibfnamefont {P.~S.}\
  \bibnamefont {Julienne}}, \bibinfo {author} {\bibfnamefont {J.}~\bibnamefont
  {Ye}}, \bibinfo {author} {\bibfnamefont {P.}~\bibnamefont {Zoller}}, \bibinfo
  {author} {\bibfnamefont {E.}~\bibnamefont {Demler}}, \bibinfo {author}
  {\bibfnamefont {M.~D.}\ \bibnamefont {Lukin}}, \ and\ \bibinfo {author}
  {\bibfnamefont {A.~M.}\ \bibnamefont {Rey}},\ }\href@noop {} {\bibfield
  {journal} {\bibinfo  {journal} {Nature Physics}\ }\textbf {\bibinfo {volume}
  {6}},\ \bibinfo {pages} {289} (\bibinfo {year} {2010})}\BibitemShut {NoStop}%
\bibitem [{\citenamefont {Zhang}\ \emph {et~al.}(2014)\citenamefont {Zhang},
  \citenamefont {Bishof}, \citenamefont {Bromley}, \citenamefont {Kraus},
  \citenamefont {Safronova}, \citenamefont {Zoller}, \citenamefont {Rey},\ and\
  \citenamefont {Ye}}]{Zhang14}%
  \BibitemOpen
  \bibfield  {author} {\bibinfo {author} {\bibfnamefont {X.}~\bibnamefont
  {Zhang}}, \bibinfo {author} {\bibfnamefont {M.}~\bibnamefont {Bishof}},
  \bibinfo {author} {\bibfnamefont {S.~L.}\ \bibnamefont {Bromley}}, \bibinfo
  {author} {\bibfnamefont {C.~V.}\ \bibnamefont {Kraus}}, \bibinfo {author}
  {\bibfnamefont {M.~S.}\ \bibnamefont {Safronova}}, \bibinfo {author}
  {\bibfnamefont {P.}~\bibnamefont {Zoller}}, \bibinfo {author} {\bibfnamefont
  {A.~M.}\ \bibnamefont {Rey}}, \ and\ \bibinfo {author} {\bibfnamefont
  {J.}~\bibnamefont {Ye}},\ }\href {\doibase 10.1126/science.1254978}
  {\bibfield  {journal} {\bibinfo  {journal} {Science}\ }\textbf {\bibinfo
  {volume} {345}},\ \bibinfo {pages} {1467} (\bibinfo {year}
  {2014})}\BibitemShut {NoStop}%
\bibitem [{\citenamefont {Fisher}\ and\ \citenamefont
  {Hohenberg}(1988{\natexlab{b}})}]{Fisher1988}%
  \BibitemOpen
  \bibfield  {author} {\bibinfo {author} {\bibfnamefont {D.~S.}\ \bibnamefont
  {Fisher}}\ and\ \bibinfo {author} {\bibfnamefont {P.~C.}\ \bibnamefont
  {Hohenberg}},\ }\href {\doibase 10.1103/PhysRevB.37.4936} {\bibfield
  {journal} {\bibinfo  {journal} {Phys. Rev. B}\ }\textbf {\bibinfo {volume}
  {37}},\ \bibinfo {pages} {4936} (\bibinfo {year}
  {1988}{\natexlab{b}})}\BibitemShut {NoStop}%
\bibitem [{\citenamefont {Berges}(2004)}]{Berges2004}%
  \BibitemOpen
  \bibfield  {author} {\bibinfo {author} {\bibfnamefont {J.}~\bibnamefont
  {Berges}},\ }\href {\doibase http://dx.doi.org/10.1063/1.1843591} {\bibfield
  {journal} {\bibinfo  {journal} {AIP Conference Proceedings}\ }\textbf
  {\bibinfo {volume} {739}},\ \bibinfo {pages} {3} (\bibinfo {year}
  {2004})}\BibitemShut {NoStop}%
\bibitem [{\citenamefont {Moeckel}\ and\ \citenamefont
  {Kehrein}(2008{\natexlab{b}})}]{Moeckel2008}%
  \BibitemOpen
  \bibfield  {author} {\bibinfo {author} {\bibfnamefont {M.}~\bibnamefont
  {Moeckel}}\ and\ \bibinfo {author} {\bibfnamefont {S.}~\bibnamefont
  {Kehrein}},\ }\href {\doibase 10.1103/PhysRevLett.100.175702} {\bibfield
  {journal} {\bibinfo  {journal} {Phys. Rev. Lett.}\ }\textbf {\bibinfo
  {volume} {100}},\ \bibinfo {pages} {175702} (\bibinfo {year}
  {2008}{\natexlab{b}})}\BibitemShut {NoStop}%
\end{thebibliography}%

\end{document}